\documentclass[10pt]{article}
\usepackage{amssymb}
\usepackage{amsmath}
\usepackage{amsthm}
\usepackage{latexsym}
\usepackage[dvips]{epsfig}
\usepackage{mathrsfs}
\usepackage{eufrak}

\theoremstyle{plain}

\newtheorem{theorem}{Theorem}

\font\SYM=msbm10
\newcommand{\Real}{\mbox{\SYM R}}

\newcommand{\Natural}{\mbox{\SYM N}}

\newcommand{\Sphere}{\mbox{\SYM S}}

\font\tenscr=rsfs10 scaled1100
\font\sevenscr=rsfs7 
\font\fivescr=rsfs5 
\skewchar\tenscr='177
\skewchar\sevenscr='177
\skewchar\fivescr='177
\newfam\scrfam
\textfont\scrfam=\tenscr
\scriptfont\scrfam=\sevenscr
\scriptscriptfont\scrfam=\fivescr

\def\scri{{\fam\scrfam I}}

\newcommand{\updn}[3]{#1^{#2}_{\phantom{#2}#3}}
\newcommand{\dnup}[3]{#1_{#2}^{\phantom{#2}#3}}

\newcommand{\dnupdn}[4]{#1_{#2 \phantom{#3} #4}^{\phantom{#2}#3}}

\newcommand{\tensor}[3]{_{#1 \phantom{#2}#3}^{\phantom{#1}#2}}
\def\o{o}
\def\i{\iota}
\def\Ad{{A'}}
\def\Bd{{B'}}
\def\Cd{{C'}}
\def\Dd{{D'}}
\def\Ed{{E'}}
\def\Fd{{F'}}

\def\es{{\bar{s}}}
\def\er{{\bar{r}}}

\newcounter{mnote}

\begin{document}



\title{\textbf{On de Sitter-like and Minkowski-like space-times}}

\author{
{\Large Christian L\"ubbe} \thanks{E-mail address:
 {\tt c.luebbe@qmul.ac.uk}} \\
{\Large Juan Antonio Valiente Kroon} \thanks{E-mail address:
 {\tt j.a.valiente-kroon@qmul.ac.uk}} \\
 \emph{Queen Mary University of London}, \\
 \emph{School of Mathematical Sciences},\\
\emph{Mile End Road, London E1 4NS, United Kingdom.}}

\maketitle

\begin{abstract}
  Friedrich's proofs for the global existence results of de
  Sitter-like space-times and of semi-global existence of
  Minkowski-like space-times [Comm. Math. Phys. \textbf{107}, 587
  (1986)] are re-examined and discussed, making use of the extended
  conformal field equations and a gauge based on conformal geodesics.
  In this gauge the location of the conformal boundary of the
  space-times is known \emph{a priori} once the initial data has been
  prescribed. Thus it provides an analysis which is conceptually and
  calculationally simpler.
\end{abstract}


\section{Introduction}
In \cite{Fri86b, Fri86c} the existence and stability of vacuum de
Sitter-like space-times has been discussed. Moreover, reference
\cite{Fri86b} provides a semi-global existence and stability result
for the development of hyperboloidal initial data which is close to
Minkowski data. These results were subsequently generalised to the
case where the gravitational field is coupled to Maxwell and
Yang-Mills fields in \cite{Fri91}. These results make use of the
Einstein conformal field equations---see e.g.
\cite{Fri81,Fri81a,Fri81b,Fri83}--- to reformulate Cauchy problems
which are global or semi-global in time into problems which are local in time.
Given one of these local Cauchy problems, then using powerful results of the theory of quasi-linear symmetric
hyperbolic systems, e.g. \cite{Kat70,Kat73,Kat75}, it is possible to
prove the existence of solutions which are close to some explicitly
known reference solutions ---the de Sitter space-time and the Minkowski
space-time. This particular strategy to prove global and semi-global
existence and stability only works in 4 dimensions, although it
should be mentioned that alternative proofs \cite{And04, AndChr05} have been obtained using the
so-called Fefferman-Graham conformal invariants which are valid in
arbitrary even dimensions.

When discussing the conformal structure of space-times it can prove
valuable to make use of gauges based on conformal invariants. One of
these invariants, the conformal geodesics, has been introduced in
\cite{FriSch87} as a tool for the local analysis of the structure of
conformally rescaled space-times. These conformal geodesics are
associated to conformal structures in a similar way as geodesics are
related to a metric. More importantly, conformal geodesics retain
their character upon conformal rescalings. As in the case of the usual
Gaussian coordinates, coordinates on a fiduciary spacelike
hypersurface are kept constant along a fixed congruence of timelike
conformal geodesics. In \cite{Fri03c} it has been shown that on the
Schwarzschild-Kruskal space-time it is possible to construct a system
of globally defined conformal Gaussian coordinates.

In \cite{Fri95} a more general set of conformal field equations has
been derived: \emph{the extended conformal Einstein field equations}.
These conformal equations are expressed using a so-called \emph{Weyl
  or conformal connection}. A Weyl connection is a torsion free
connection (not necessarily Levi-Civita) which preserves the conformal
 metric and for which parallel transport preserves conformally orthonormal
 frames and the causal nature of their vectors ---that is, time-like vectors
are transported into time-like vectors, etc. The extra freedom
introduced by the use of Weyl connections allows one to consider gauges
based on conformal geodesics in the discussion of various local and
global issues in General Relativity. For example, in \cite{Fri95} this
type of gauge and the evolution equations implied by them have been
used to discuss the (local) existence of anti de Sitter-like
space-times; in \cite{Fri98a} conformal geodesics have been used to
construct a representation of spatial infinity which makes possible,
in principle, the detailed discussion of the structure of the
gravitational field in this region of space-time.

\medskip One of the properties of conformal Gaussian
coordinates is that in the case of vacuum space-times, they provide a
canonical conformal factor, written entirely in terms of quantities
defined on a fiduciary hypersurface and the conformal time parameter.
In turn, this conformal factor provides \emph{a priori} knowledge of 
the structure of the conformal boundary of the space-time. In this article 
we make use of this aspect to re-examine and discuss the
global and semi-global existence results of \cite{Fri86b}. The
possibility of following this approach is somehow implicit in the
literature, but to our knowledge has never been made explicit.
The approach discussed here is in a sense more natural as it exploits 
the full freedom contained in the extended conformal field
equations by making use of a gauge based on conformal invariants. In
the original approach used in \cite{Fri86b} the conformal factor is
itself an unknown satisfying a propagation equation. Consequently the
properties of the conformal boundary of the resulting space-time have
to be discussed \emph{a posteriori} in an abstract way. For example,
in the case of the development of hyperboloidal data, the existence of
a point corresponding to future time-like infinity where the
generators of null infinity meet is inferred from a qualitative
argument involving the Raychauduri equations. In contrast, in
the approach followed here the existence and properties of future
time-like infinity follow from explicit calculations and the
requirement that the data is close to Minkowski data. Moreover, this
information is available without reference to the existence problem of
the propagation equations.

\subsection{Structure and overview of the article}
This article is structured as follows. Section \ref{section:setup}
discusses congruences of conformal geodesics in vacuum space-times and
the construction of conformal Gaussian coordinates on space-times
whose time slices are diffeomorphic to the 3-sphere, or subsets
thereof. The canonical conformal factors associated to the congruence
of conformal geodesics are also studied. Section
\ref{asimple:model:solns} is concerned with the conformal de Sitter
and Minkowski space-times as subsets of the Einstein cylinder and with
congruences of conformal geodesics on these manifolds. These
space-times are analysed in detail as they will be used as reference
solutions later on.  Section \ref{section:conformal:boundary}
discusses the conformal boundary of the development of initial data
sets which are sufficiently close to Cauchy data for the de Sitter
space-time or to hyperboloidal data for the Minkowski space-time.
Section \ref{section:evolution:eqns} is concerned with the propagation
equations implied by the extended conformal equations and the
conformal Gaussian coordinates. The propagation equations are
expressed in terms of a space spinor formalism. In section
\ref{reference:st:as:solns} the de Sitter and Minkowski space-times
are recast as solutions of the propagation equations of section
\ref{section:evolution:eqns}. Section \ref{section:existence:results}
discusses the global and semi-global existence results implied by the
propagation equations of section \ref{section:evolution:eqns}. These
existence results are a consequence of a modified version of an
existence and stability result for quasilinear hyperbolic systems by
T. Kato. This modified Kato theorem was first discussed in
\cite{Fri86b} and is included here for completeness (theorem
\ref{thm:mod:Kato}). The existence results considered here are: global
existence for de Sitter-like space-times when data are given either on
a standard Cauchy hypersurface (theorem \ref{thm:standard:deSitter})
or on the conformal boundary (theorem \ref{thm:asymptotic:deSitter});
and semi-global existence for hyperboloidal Minkowski-like data
(theorem \ref{thm:Minkowski}).  Section \ref{section:conclusions}
provides some concluding remarks about possible generalisations of the
results provided in the present work. An alternative discussion of the
behaviour of conjugate points in the congruence of conformal geodesics
is provided in the appendix.

\section{Setup and gauge considerations}
\label{section:setup}
Let $(\tilde{\mathcal{M}},\tilde{g}_{\mu\nu})$ be a space-time
satisfying the Einstein field equations with cosmological constant
\[
\tilde{R}_{\mu\nu}=\lambda \tilde{g}_{\mu\nu}.
\]
We will only be concerned with the case $\lambda\leq 0$.
Space-times such that $\lambda<0$ and suitably close to the de
Sitter solution will be called \emph{de Sitter-like} whereas if
$\lambda=0$ and the space-time is suitably close to Minkowski
space-time it will be called \emph{Minkowski-like}. The metric
$\tilde{g}_{\mu\nu}$ is assumed to have signature $(+,-,-,-)$.

\subsection{Coordinatising manifolds diffeomorphic to $\Sphere^3$}
We shall work with space-times which are of the form $I\times
\mathcal{S}$ where $I$ is an interval on $\Real$ and $\mathcal{S}$ is
diffeomorphic to $\Sphere^3$ or to a submanifold thereof. The manifold
$\Sphere^3$ will always be thought of as the following submanifold of
$\Real^4$:
\[
\Sphere^3 =\left \{ x^\mathcal{A} \in \Real^4 \;\bigg | \; (x^1)^2 + (x^2)^2 + (x^3)^2 + (x^4)^4=1  \right\}.
\]
The restrictions of the functions $x^\mathcal{A}$,
$\mathcal{A}=1,2,3,4$ on $\Real^4$ to $\Sphere^3$ will again be
denoted by $x^\mathcal{A}$. The vector fields
\begin{subequations}
\begin{eqnarray}
&& c_1 \equiv x^1 \partial_4 -x^4\partial_1 +x^2\partial_3 -x^3\partial_2, \label{v_field_c1} \\
&& c_2 \equiv x^1 \partial_3 -x^3\partial_1 +x^4\partial_2 -x^2\partial_4, \label{v_field_c2}\\
&& c_3 \equiv x^1 \partial_2 -x^2\partial_1 +x^3\partial_4 -x^4\partial_3. \label{v_field_c3}
\end{eqnarray}
\end{subequations}
on $\Real^4$ are tangent to $\Sphere^3$. In the sequel they will
always be considered as vectors on $\Sphere^3$.  Denote by
$\mbox{d}\omega^2$ the line element obtained as pull-back of
\[
\sum_{\mathcal{A}=1}^4 (\mbox{d} x^\mathcal{A})^2
\]
to $\Sphere^3$. The vector fields $c_1$, $c_2$, $c_3$ constitute a
globally defined frame on $\Sphere^3$ which is orthonormal with
respect to $\mbox{d}\omega^2$.

\bigskip Let $\phi: \mathcal{S} \rightarrow \Sphere^3$ denote the diffeomorphism connecting
$\mathcal{S}$ and $\Sphere^3$. The diffeomorphism $\phi$ will be employed to pull-back the
functions $x^\mathcal{A}$, $\mathcal{A}=1,\,2,\,3,\,4$ on $\Sphere^3$
to $\mathcal{S}$ . Their pull-back to $\mathcal{S}$ will again be
denoted by the same symbol. This system of equations has rank 4 on
$\mathcal{S}$, so that one can use a suitable choice of three of the
functions $x^\mathcal{A}$, to obtain a coordinate system in a
neighbourhood of a point in $\mathcal{S}$. Moreover, given
$x^\mathcal{A}$ on $\mathcal{S}$, one can define the vectors $c_1,\;
c_2,\;c_3$ as given by (\ref{v_field_c1})-(\ref{v_field_c3}). In the
following we will extend this frame by $c_0$ to a frame in $ \mathcal{M}$.
The index letters $\es=0,1,2,3$, respectively $\er = 1,2,3$, will be
specifically reserved to denote components with respect to this frame.

\subsection{Weyl connections and conformal geodesics}
Let
\[
g_{\mu\nu}= \Theta^2 \tilde{g}_{\mu\nu}
\]
be a conformally related metric where $\Theta$ is some conformal
factor. Let $b_{\mu}$ be a smooth 1-form. Denote by $\tilde{\nabla}$
and $\nabla$ the Levi-Civita connections of $\tilde{g}$ and $g$,
respectively, and by $\hat{\nabla}$ the Weyl connection for
$\tilde{g}$ satisfying $\hat{\nabla}=\tilde{\nabla} + S(b)$ , where
\[
\dnupdn{S(b)}{\mu}{\rho}{\nu}\equiv \updn{\delta}{\rho}{\mu} b_\nu + \updn{\delta}{\rho}{\nu}b_\mu -g_{\mu\nu} g^{\rho\lambda}b_\lambda.
\]
Then $\nabla = \tilde{\nabla} + S(\Upsilon) $ with $\Upsilon =
\Theta^{-1} \nabla \Theta$ and $\hat{\nabla} = \nabla + S(f)$ with $f=
b-\Upsilon$.  Further, define $d_\mu = \Theta b_\mu = \Theta f_\mu
+\nabla_\mu \Theta$. 
The Schouten tensor associated to the Weyl connection $\hat{\nabla} $ is given by
\[
\hat{L}_{\mu\nu} \equiv \frac{1}{2}\left( \hat{R}_{(\mu\nu)} -\frac{1}{2}\hat{R}_{[\mu\nu]} -\frac{1}{6} \hat{R}_{\lambda\rho} g^{\lambda\rho} g_{\mu\nu} \right) .
\]
We note that whenever the connection preserves a metric in the conformal class, i.e. it is a Levi-Civita connection, then $\hat{L}_{[\mu\nu]} = 0 = \hat{R}_{[\mu\nu]} $.

Let $e_k$, $k=0,\ldots,3$, be a $g$-orthonormal
frame field, i.e. satisfying $g(e_i,e_j)=\eta_{ij}$, with
$\eta_{ij}\equiv\mbox{diag}(1,-1,-1,-1)$, $i,j=0,\ldots,3$. Denote by
$\nabla_k$ and $\hat{\nabla}_k$ the covariant derivative in the
direction of $e_k$ with respect to $\nabla$ and $\hat{\nabla}$. Define
the connection coefficients $\dnupdn{\hat{\Gamma}}{i}{j}{k}$ of
$\hat{\nabla}$ in this frame by $\hat{\nabla}_i e_k =
\dnupdn{\hat{\Gamma}}{i}{j}{k} e_j$. A \emph{conformal geodesic}
$x(\tau)$ is obtained, together with a 1-form $b(\tau)$ along the
curve, as a solution to the system of equations
\begin{eqnarray}
&& \dot{x}^\nu \tilde{\nabla}_\nu \dot{x}^\mu + \dnupdn{S(b)}{\lambda}{\mu}{\rho} \dot{x}^\lambda \dot{x}^\rho =0,\nonumber \\
&& \dot{x}^\nu \tilde{\nabla}_\nu b_\mu -\frac{1}{2} b_\nu \dnupdn{S(b)}{\lambda}{\nu}{\mu} \dot{x}^\lambda = \tilde{L}_{\lambda\mu} \dot{x}^\lambda,\nonumber
\end{eqnarray}
where $\dot{x}$ denotes the tangent vector to the curve $x(\tau)$ and
$\tilde{L}_{\mu\nu} = \tilde{\lambda} \tilde{g}_{\mu\nu} $ with $\lambda = 6
\tilde{\lambda}$.  In what follows we shall often write $v^\mu$ for
$\dot{x}^\mu$.  Given initial data for these equations in the form
$x_*\in \tilde{\mathcal{M}}$, $\dot{x}_*\in T_{x_*}
\tilde{\mathcal{M}}$, $b_* \in T^*_{x_*} \tilde{\mathcal{M}}$, there
exists a unique conformal geodesic $(x(\tau)$, $b(\tau))$ near $x_*$
satisfying for given $\tau_0 \in \Real$
\[
x(\tau_0)=x_*, \quad \dot{x}(\tau_0)=\dot{x}_*, \quad b(\tau_0)=b_*.
\]
Conformal geodesics are conformally invariant in the sense that if
$x(\tau)$ and $b(\tau)$ solve the conformal geodesics equations and we
define a new Weyl connection $\check{\nabla}=\tilde{\nabla}+
S(\check{b})$, then $(x(\tau),b(\tau)-\check{b}(\tau))$ solve the
conformal geodesic equations with $\tilde{\nabla}$ replaced by
$\check{\nabla}$ and $\tilde{L}_{\mu\nu}$ by $\check{L}_{\mu\nu}$. In
particular if $\check{b} = b$ then $\hat{\nabla}$ and $\check{\nabla}
$ coincide and the conformal geodesic equations take the form
\begin{eqnarray}
&& \dot{x}^\nu \hat{\nabla}_\nu \dot{x}^\mu=0, \nonumber\\
&& \hat{L}_{\mu\nu} \dot{x}^\mu =0.\nonumber
\end{eqnarray}

\subsection{Conformal  Gaussian coordinates}
Let $\tilde{\mathcal{S}}$ be a spacelike hypersurface in the vacuum
space-time $(\tilde{\mathcal{M}},\tilde{g})$. Let
$\tilde{h}_{\alpha\beta}$ denote the intrinsic 3-metric of
$\tilde{\mathcal{S}}$ induced by $\tilde{g}_{\mu\nu}$. On
$\tilde{\mathcal{S}}$ choose:
\begin{itemize}
\item[(i)] a positive conformal factor $\Theta_*$;
\item[(ii)] a frame field $e_{*k}$, $k=0,\ldots,3$  such that
$\tilde{g}(e_{*i},e_{*k})=\Theta^{-2}_*\eta_{ik}$, 
\item[(iii)] and a 1-form $b_*$.
\end{itemize}
Given the above initial information, there exists through each point
$x_*\in \tilde{\mathcal{S}}$ a unique conformal geodesic $(x(\tau)$,
$b(\tau))$ with $\tau=\tau_*$ on $\tilde{\mathcal{S}}$ which satisfies
the initial conditions $\dot{x}(\tau_*)=e_{0*}$, $b(\tau_*)=b_*$.
These curves define a smooth-caustic free congruence in a
neighbourhood $\mathcal{U}$ of $\mathcal{S}$ if all data are smooth.
In addition, $b$ defines a 1-form on $\mathcal{U}$ from which one can
construct a Weyl connection $\hat{\nabla}=\tilde{\nabla} + S(b)$. A
smooth frame field $e_k$ and a conformal factor $\Theta$ are then
obtained on $\mathcal{U}$ by solving the propagation equation
\[
\dot{x}^\mu \hat{\nabla}_\mu e_k =0.
\]
It can be seen that $\tilde{g}(e_i,e_j)=\Theta^{-2} \eta_{ij}$ on $\mathcal{U}$ with
\begin{equation}
\label{conformal factor}
\dot{x}^\mu \hat{\nabla}_{\mu} \Theta = \Theta \langle b, \dot{x} \rangle, \quad \Theta|_{\mathcal{S}}=\Theta_*,
\end{equation}
where $\langle \cdot, \cdot \rangle$ denotes the contraction of a
1-form with a vector. Accordingly, the frame $e_k$ is orthonormal for
the metric $g_{\mu\nu}=\Theta^2\tilde{g}_{\mu\nu}$. Coordinates
$x^\mathcal{A}$ on $\tilde{\mathcal{S}}$ can be dragged along the
congruence of conformal geodesics, so that if one sets $x^0=\tau$, one
obtains a coordinatisation of $\mathcal{U}\subset
\tilde{\mathcal{M}}$. In this gauge one has that
\begin{equation}
\label{cggauge}
\dot{x} = v = e_0 = \partial_\tau, \quad \dnupdn{\hat{\Gamma}}{0}{j}{k}=0, \quad \hat{L}_{0k}=0.
\end{equation}
Such a choice of coordinates, frame field and conformal gauge will be
referred to as a \emph{conformal Gaussian system}. Let $g^{\mu\nu}=
h^{\mu\nu}+ v^\mu v^\nu$ where the pull-back of $h^{\mu\nu}$ is the
contravariant negative definite intrinsic metric of the surfaces
orthogonal to the congruence. The Levi-Civita connection of
$h_{\mu\nu}$ will be denoted by $D$. Since $v = e_0$ is orthogonal to
$\tilde{\mathcal{S}}$, $h_{\mu\nu}$ $\mu,\nu=0,\ldots3$, coincides
with the 3-metric $h_{\alpha\beta}\equiv \Theta_*^2
\tilde{h}_{\alpha\beta}$ $\alpha,\beta=1,2,3$. In general it will
however not agree with the 3-metric of the surface of
$\tau=\mbox{constant}$.

\bigskip
If $\tilde{g}$ is a solution to the Einstein vacuum field
equations with cosmological constant $\lambda$ then for a conformal Gaussian system the conformal factor
$\Theta$ and the 1-form $d_k$ can be determined explicitly from equation (\ref{conformal factor}) and the initial data. 
More precisely, one has that
\begin{subequations}
\begin{eqnarray}
\Theta_* \ne 0& & \Theta = \Theta_* \left( 1 + \tau \langle b_*, \dot{x}_* \rangle + \frac{\tau^2}{2} \left( \tilde{\lambda} \Theta^{-2}_* + \frac{1}{2} g^\sharp(b_*,b_*) \right) \right), \label{canonical_Theta1}\\
\Theta_* =0 & & \Theta 
= \langle d_*, v_* \rangle \tau + \frac{1}{2} \ddot{\Theta}_* \tau^2,
\label{canonical_Theta0}
\end{eqnarray}
\end{subequations}
and
\begin{equation}
d_0= \dot{\Theta}, \quad d_a = \langle b_*, \Theta_* e_{a*} \rangle, \quad a=1,2,3, \label{canonical_d}
\end{equation}
where we have set $b_* = \Upsilon_* $, $\tau_*=0 $ and used the identity
\begin{equation}
\ddot{\Theta}\Theta= \tilde{\lambda}  + \frac{1}{2} g^\sharp(d,d),
\label{ddTheta}
\end{equation}
In the above expressions, quantities with the subscript ${}_*$ are
regarded as constant along the conformal geodesics, while
$g^\sharp(\cdot,\cdot)$ denotes the contravariant form of the metric
$g_{\mu\nu}$ applied to a pair of 1-forms. As long as the congruence
of conformal geodesics does not degenerate, then at the points where
$\Theta=0$, $\nabla_k \Theta\neq 0$ one finds that
\[
g^\sharp(d,d)= \eta^{kl} \nabla_k \Theta \nabla_l \Theta =-2\tilde{\lambda}.
\]
The nature of the conformal boundary (spacelike, timelike, null), defined by the conditions $\Theta=0$, $\nabla_k \Theta\neq 0$, can be deduced according to whether $\tilde{\lambda}$ is negative, positive or zero, respectively.

\subsection{Conjugate points}

As mentioned in the previous paragraphs, it is necessary to check that
the conformal geodesic congruence does not develop any conjugate
points or caustics. This would lead to a break down of the conformal
Gaussian coordinates. For our analysis we use the conformal Jacobi
fields $\eta^\mu = \eta^k e^\mu_k $, as their component $h_{jk}\eta^k
$ vanishes at a conjugate point \cite{Fri03c}.  They satisfy
\begin{subequations}
\begin{eqnarray}
&& \partial_\tau  \eta_k = \chi_{jk}\eta^j,\label{Jacobi1}\\
&& \partial_\tau^2 \eta_k = -( E_{jk} + \hat{L}_{jk})\eta^j, \label{Jacobi2} \\
&& \partial_\tau^3 \eta_k = -\partial_\tau(E_{jk})\eta^j) + \hat{Y}_{0jk}\eta^j, \label{Jacobi3}
\end{eqnarray}
\end{subequations}
where $\chi_{jk}$ denotes the components, with respect to the frame
$e_k$, of the second fundamental form of the surfaces of constant
$\tau$, and we have used the 4-dimensional Cotton-York tensor
$\hat{Y}_{ijk}\equiv\hat{\nabla}_{[i}\hat{L}_{j]k}$, and the electric
part of the Weyl tensor, $E_{ij}\equiv C_{0i0j}$.

\section{The asymptotically simple model solutions}
\label{asimple:model:solns}

In this section we discuss the exact solutions of de Sitter
($\tilde{\lambda}= -1/2 $) and Minkowski ($\tilde{\lambda}= 0 $) in
terms of conformal geodesics and initial data for the regular
conformal field equations. It is well known \cite{HawEll73} that both
solutions can be conformally embedded into the Einstein cylinder
$\Real \times \Sphere^3 $ 
with metric
\[
g_E = \mbox{d}^2 t - \mbox{d}^2\chi - \sin^2 \chi \mbox{d}\sigma^2,
\]
where $\chi\in[0,\pi]$ and $\mbox{d}\sigma$ denotes the standard
metric of $\Sphere^2$. The curves $x^\mu(t) = x^\mu_* + t \delta^\mu_0
= x^\mu_* + t u^\mu $ are conformal geodesics written in terms of the
coordinate $t$ and $u=\partial_t$. Along these curves we have
\begin{eqnarray}
&&\Theta_E(\tau) =  \left( 1 + \frac{\tau^2}{4} \right), \nonumber\\
&&v(\tau) = \Theta_E(\tau)^{-1} u, \nonumber\\
&&b_E(\tau) =\frac{\tau}{2} \, \mbox{d}t = \frac{\mbox{d} \Theta_E}{\Theta_E},\nonumber
\end{eqnarray}
with
\[
\tau \equiv 2\tan \frac{t}{2}, 
\]
and where $\Theta_E $ is the conformal factor generated along the
geodesic by (\ref{conformal factor}) with $\Theta_{E*}=1$ at
$t=0$. Observe that $\tau \to \pm \infty$, $\Theta_E \to \infty$ and
$v \to 0$ as $t \to \pm \pi$.  The metric $g\equiv \Theta_E^2 g_E$ is
given by
\begin{equation}
\label{g_E}
g= \Theta_E^2 g_E =\mbox{d}^2 \tau - \left( 1 + \frac{\tau^2}{4} \right) \left( \mbox{d}^2\chi + \sin^2 \chi \mbox{d}\sigma^2 \right)
\end{equation}
and satisfies $g(v,v)=1$. Note that all curves are orthogonal to the
surfaces of constant $\tau$. In the rescaled space-time
$b=b_E-\Theta_E^{-1}\mbox{d} \Theta_E=0$, and thus, the curves are
geodesics with respect to $g $.

\subsection{The de Sitter space-time}
\label{section:deSitter}
The de Sitter space-time is embedded into the Einstein cylinder using
the conformal factor
\[
\Omega_D = \cos t = \left( \frac{ 1-\frac{\tau^2}{4} }{1+\frac{\tau^2}{4}}\right) = \frac{2-\Theta_E}{\Theta_E}.
\]
The conformal factor $\Omega_D$ vanishes at $\tau = \pm 2$, where the
conformal boundary $\mathscr{I}=\mathscr{I}^-\cup \mathscr{I}^+$ is located.
Using
\[
\Upsilon_D = \frac{d \Omega_D}{\Omega_D} =
-\frac{\tau}{1-\frac{1}{4}\tau^2} \, \mbox{d}t,
\]
setting $\Theta_{D*}=1$ and using equation (\ref{conformal factor}) we get that along the congruence of conformal geodesics
\begin{eqnarray}
&&\Theta_D(\tau)= 1-\frac{1}{4}\tau^2 = \Omega_D \Theta_E, \nonumber\\
&& b_D = b_E + \Upsilon_D = \frac{-\frac{1}{2}\tau}{1-\frac{1}{4}\tau^2} \, \mbox{d}\tau = \frac{\mbox{d}(\Omega_D \Theta_E)}{\Omega_D \Theta_E}.\nonumber
\end{eqnarray}
On the Cauchy surface $\tau=0$ we have $b_D(0)=0 $. From equation
(\ref{canonical_Theta1}) we recover $\Theta_D $, as above.  We see
that rescaling by $\Theta_D$ gives the metric $g$ once more. The
Cauchy surface $\tau=-2$ represents the past conformal boundary,
$\mathscr{I}^-$. If we redefine $\tau \to \tau + 2 $ we get on the
conformal de Sitter space-time that
\begin{eqnarray}
&&\Theta_D = \tau - \frac{1}{4}\tau^2, \nonumber\\
&&d_D = \Theta_D b_D = - \frac{1}{2} (\tau-2) \mbox{d}\tau, \nonumber\\
&&v=\frac{u}{1 + \frac{1}{4}(\tau-2)^2},\nonumber
\end{eqnarray}
with the initial data at $\mathscr{I}^-$ given by 
\[
d_{D*} = \mbox{d}\tau
,\quad \langle d_D, v\rangle_* = 1 , \quad \ddot{\Theta}_{D*}= -
\frac{1}{2}.
\]
We observe that in this case $\Theta_D$ is given by the formula (\ref{canonical_Theta0}).

\subsection{Minkowski space-time}
\label{section:Minkowski:reference}

The Minkowski space-time will be embedded using the conformal factor
\[
\Omega_M = \cos(t+\frac{\pi}{2}) + \cos \chi .
\]
For convenience we have shifted the standard embedding by $\pi/2$ to
the past here, so that the usual Minkowskian hyperboloid, which is
usually embedded at $t = \pi/2$, is now located at $t=0$. We have
\[
\Upsilon_M = \frac{\mbox{d} \Omega_M}{\Omega_M} = \frac{\cos t \,\mbox{d}t + \sin \chi \,\mbox{d}\chi  }{ \sin t - \cos \chi  }.
\]
It follows that the conformal geodesics satisfy
\begin{eqnarray}
&&b_M = b_E + \Upsilon_M = \tan\left(\frac{t}{2}\right) \mbox{d}t + \frac{\cos t \,\mbox{d}t + \sin \chi \,\mbox{d}\chi  }{ \sin t - \cos \chi  }, \nonumber \\
&&v = \cos^2\left(\frac{t}{2}\right) u, \nonumber
\end{eqnarray}
Hence at $t=0=\tau$, we get the following information on the canonical
Minkowski hyperboloid
\begin{eqnarray}
&&\Omega_{M*} = \cos \chi, \nonumber\\
&& d_{M*}= -(\mbox{d}t + \sin \chi), \nonumber\\
&& \langle d_M, v \rangle_* =-1, \nonumber\\
&& h^\sharp(d,d)_* = \sin^2 \chi .\nonumber
\end{eqnarray}
Substituting these into formula (\ref{canonical_Theta1}) we get
$$\Theta_M = \cos \chi \left( 1 - \frac{\tau}{\cos \chi} + \frac{\tau^2}{4} \right) = \Omega_D \Theta_E ,$$
which vanishes at
\[
\tau = 2\frac{1 \pm \sin \chi}{\cos \chi}.
\]

\subsection{Conjugate points in the reference solutions}
For the reference solutions discussed in this section the electric
part of the Weyl tensor $E_{jk}$ and the 4-dimensional Cotton-York
tensor $\hat{Y}_{ijk}$ vanish. Furthermore, the components of the
second fundamental form $\chi_{jk}(0)$ vanishes. Using equations
(\ref{Jacobi1})-(\ref{Jacobi3}) one finds for the chosen congruence
\[
\eta_0(\tau)=\eta_0(0), \quad\eta_k(\tau) = \Theta_E(\tau)\eta_k(0).
\]
Now, $\Theta_E \neq 0$ for $\tau\in(-\infty,\infty)$.  Thus, the Jacobi
fields will never be tangent to the curve nor vanish.  Hence the
congruence is free of conjugate points.

\section{The structure of the conformal boundary for nearby
  space-times}
\label{section:conformal:boundary}
In this section we use the formulae (\ref{canonical_Theta1}) and
(\ref{canonical_Theta0}) to study the conformal boundary of
space-times which are constructed as the development of initial data
which is close to either de Sitter Cauchy data or to hyperboloidal
Minkowski initial data.

\subsection{Space-times close to de Sitter}
As discussed in \cite{Fri86c}, for de Sitter-like space-times one can
formulate two slightly different Cauchy initial problems: one where
data is prescribed on a standard Cauchy hypersurface, and a second one
where the data is prescribed precisely on one portion of the conformal
boundary, $\mathscr{I}^-$.

\subsubsection{The case of de Sitter-like data away from $\mathscr{I}^- $ }
\label{section:standard:deSitter:data}
Here we assume that we are given a space-like hypersurface
$\mathcal{S}$ which does not intersect $\mathscr{I}^-$. So $\Theta$ is
given by (\ref{canonical_Theta1}) as $\Theta_* \ne 0 $ on
$\mathcal{S}$. In fact, without loss of generality we could set
$\Theta_*=1 $. The conformal factor vanishes at
\begin{equation}
\label{taupmdS1}
\tau_\pm = \frac{ - 2\Theta_*\langle d, v \rangle_* \pm 2\Theta_* \sqrt{| 2\tilde{\lambda} + g^\sharp(d,d)_*  |} }{ 2\tilde{\lambda} + g^\sharp(d,d)_* }
\end{equation}
which gives the location of the conformal boundary, $\mathscr{I}^\pm$
as smooth space-like hypersurfaces. If
$\mathcal{S}$ is topologically $\Sphere^3$ ---as it is being assumed
here--- then
\[
\mathscr{I}^\pm =\{ \tau_\pm \} \times \mathcal{S}
\]
are topologically $\Sphere^3$.  On $\mathscr{I}^\pm $ we have
$\nabla_k \Theta \nabla^k \Theta =-2\tilde{\lambda}$ and hence both
hypersurfaces are space-like.

\subsubsection{The case of de Sitter-like data on $\mathscr{I}^-$}
\label{section:asymptotic:deSitter:data}
We discuss now the data for a conformal geodesic congruence that
starts on the smooth hypersurface $\mathcal{S}$ which represents
$\mathscr{I}^-$.  Hence, the initial data satisfies the condition
$\Theta_*=0 $ on $\mathcal{S}$ and takes the form
(\ref{canonical_Theta0}). Equation (\ref{ddTheta}) implies that
$g^\sharp(d,d)_*=-2 \tilde{\lambda}$ and thus $d_*$ must be time-like
as $ \tilde{\lambda}<0$. On the other hand, $\ddot{\Theta}_* $  is free
data on $\mathcal{S}$. Having set $b_* = \Upsilon_* $, that is, $d_* =
(\nabla \Theta)_*$ and $v_*=n$, where $n$ is the unit normal of
$\mathcal{S}$ with respect to $g$, it follows from $\Theta_* =0$ on
$\mathcal{S}$ that
\[
d_*= \langle d, v \rangle_* v_* \quad \mathrm{with} \quad \dot{\Theta}_*= \langle d, v \rangle_* = \pm \sqrt{-2\tilde{\lambda}} .
\]
We choose the positive root so that $\Theta $ is positive in the
future of $\mathcal{S}$. Thus with respect to the Weyl propagated
orthonormal frame $e_k $ we obtain from \ref{canonical_d}
\[
d_k(\tau)=\left(\sqrt{-2\tilde{\lambda}} + \ddot{\Theta}_* \tau, 0, 0, 0\right) .
\]
The conformal factor vanishes at
\begin{eqnarray}
&& \mathscr{I}^- =  \{ 0 \} \times \mathcal{S}, \nonumber \\
&& \mathscr{I}^+ = \left\{\tau= -2\frac{\dot{\Theta}_*}{\ddot{\Theta}_*} \right\}\times \mathcal{S}. \nonumber
\end{eqnarray}
Hence, the location of $\mathscr{I}^+ $ is determined by the free data
$ \ddot{\Theta}_*$. On $\mathscr{I}^+ $ we have $d(\tau_+)=-d_* $ and
$\nabla_k \Theta \nabla^k \Theta =-2\tilde{\lambda}$ and hence, again,
it is a space-like hypersurface.

\subsection{Hyperboloidal Minkowski data}
\label{section:Minkowski:data}

We discuss now how to use formula (\ref{canonical_Theta1}) to gain \emph{a
priori} information on the conformal boundary of the domain of
dependence of hyperboloidal initial data which is close to Minkowski
data. Given a 3-dimensional manifold $\mathcal{S}$ with the topology of $\Sphere^3$, we
consider $\overline{\mathcal{S}}\subset \mathcal{S}$, with
$\partial \overline{\mathcal{S}}$ diffeomorphic to $\Sphere^2$. Furthermore, consider a function $\Omega$ on
$\mathcal{S}$ such that $\Omega>0$ in the interior of
$\overline{\mathcal{S}}$, and $\Omega=0$ on $\mathcal{Z}\equiv
\partial \overline{\mathcal{S}}$. The function $\Omega$ can be
obtained naturally as part of a solution to the conformal Hamiltonian
and momentum constraint --- for a discussion and more details on this
see \cite{Fri83,Fri86b}.

\medskip
It is noted that
\[
g^\sharp(b_*,b_*)=h^\sharp(b_*,b_*) + \langle b_*, v_* \rangle^2,
\]
where $h^\sharp(b_*,b_*) \leq 0$, and
$|h^\sharp(b_*,b_*)|=-h^\sharp(b_*,b_*)$ since $h_{\mu\nu}$ is taken
to be negative definite.  In order to make use of formula
(\ref{canonical_Theta1}), the following particular choices of initial
data will be made
\[
\Theta_*=\Omega, \quad \langle d_*,v_* \rangle = \dot{\Theta}_*, \quad d_{*a} = \Omega b_{*a} = D_a\Omega,
\]
where $b_{*a}= e^\mu_a b_{*\mu}$, $D_a = e^\mu_a \nabla_\mu=e^\alpha_a
D_\alpha$, $a=1,2,3$.  In particular, the sign of $\dot{\Theta}_*$
contains the information about which part of the locus of points such
that $\Theta=0$ should be considered as the conformal boundary ---see
the discussion below. On $\overline{\mathcal{S}} \setminus \mathcal{Z}
$ the function $ \ddot{\Theta}_*$ is determined by (\ref{ddTheta}) for
the initial data $ \ddot{\Theta}_*$ is taken to extend smoothly to
$\mathcal{Z}$. On $\mathcal{Z}$, the formula (\ref{ddTheta}) implies
that $g^\sharp(d,d)_*= g^\sharp(\nabla \Theta, \nabla \Theta)_*=0$.
For hyperboloidal data $ (\nabla \Theta)_* \ne 0 $ on $\mathcal{Z}$,
and thus $d_*$ must be null on $\mathcal{Z}$. We must thus have
\[
\Xi^2= \langle d_*,v_* \rangle^2 \quad \mbox{ on } \quad \mathcal{Z},
\]
where $\Xi\equiv \sqrt{|h^\sharp(d,d)_* |} = \sqrt{|D_k\Omega D^k
  \Omega|}$ on $\overline{\mathcal{S}} $. Consequently,
\[
h^\sharp(b_*,b_*) = -\frac{4}{\omega^2}, \quad \mbox{ with } \quad \omega \equiv \frac{2\Omega}{\Xi} .
\]
The functions $\Omega$, $\Xi$ and $\omega$ will be extended off $\mathcal{S}$
by requiring that they remain constant along a given curve of the
congruence of conformal geodesics, and will be denoted again by
$\Omega$, $\Xi$, $\omega$. 

In order to further discuss the structure of the
conformal boundary, we analyse the zeros of $\Theta$. For curves
passing through $\mathcal{Z}$ the zeros are located at
\[
\tau_\pm  = \frac{-\dot{\Theta}_* \pm \dot{\Theta}_*}{\ddot{\Theta}_*},
\]
whereas on $\overline{\mathcal{S}} \setminus \mathcal{Z}$ we can write
\begin{equation}
\Theta = \Theta_* \left( 1 + \alpha \tau + \left( \frac{1}{4}\alpha^2 -\frac{1}{\omega^2}  \right) \tau^2  \right),\label{massaged_Theta}
\end{equation}
where $\alpha\equiv \langle b_*,v_*\rangle$. Note that $\alpha$ can be
chosen independently of $\Omega$ ---this fact will play a role when
discussing the existence of solutions. The function $\alpha$ will be
extended off $\overline{\mathcal{S}}$ in the same way as it was done
for $\Omega$, $\Xi$, $\omega$. The roots of $\Theta$ are given by
\[
\tau_\pm = \frac{-2\alpha\Omega^2 \pm 2\Omega \Xi}{\alpha^2 \Omega^2 - \Xi^2  }.
\]
 Accordingly, one defines
\begin{equation}
\label{scri:Minkowski}
\mathscr{I}^\pm\equiv \bigg\{ (\tau,x^\mathcal{A}) \in \Real \times \mathcal{S} \;\bigg |\;   \tau= \tau_\pm(x^\mathcal{A})  \bigg\}.
\end{equation}
This shows that the location of $\mathscr{I}^\pm$ is predetermined by
the initial data $\Omega, d_{*0}, D_k \Omega$ and well defined as long
as the congruence does not degenerate.  One sees that as $\Theta_* \to
0 $ one has $\tau_\pm \to 0, \;-2{\dot{\Theta}_*}/{\ddot{\Theta}_*} $.
Thus, with the continuation of $\ddot{\Theta}_* $ onto $\mathcal{Z}$
described above one finds that $\mathscr{I}^+$ and $\mathscr{I}^-$ are
smooth hypersurfaces, whenever $\mathrm{d}\Theta \ne 0$, and
$\mathcal{Z}$ is the intersection of $\mathscr{I}^\pm$ with $\{0\}
\times \mathcal{S}$ as expected for hyperboloidal initial data. In
analogy to the model case of the hyperboloids in Minkowski space-time,
the development of general hyperboloidal data has a conformal boundary
which either corresponds to $\mathscr{I}^+$ or to $\mathscr{I}^-$, but
not both. This information is contained in the sign of the free datum
$\dot{\Theta}_*$.  The conformal factor is positive on the physical
space-time $(\tilde{\mathcal{M}},\tilde{g}_{\mu\nu})$. So if
$\dot{\Theta}_*>0$ on $\mathcal{Z}$, then $\tilde{\mathcal{M}} $ lies
to the future of $\mathscr{I}$. In this case one speaks of a
hyperboloid which intersects past null infinity, and thus the
conformal boundary is identified as $\mathscr{I}^-$.  Whereas if
$\dot{\Theta}_*<0$ on $\mathcal{Z}$, then $\tilde{\mathcal{M}} $ lies
to the past.  Then the hyperboloid is regarded as intersecting future
null infinity and then $\mathscr{I}^+$ gives the conformal
boundary. 
Without loss of generality, in the sequel, we shall only consider
hyperboloids intersecting future null infinity, so that
$\dot{\Theta}_*<0$ on $\mathcal{Z}$. Then $\mathscr{I}^+$ is given by $\tau_+ $ as
identified above. We remark that the solution $\tau_- $ is of no
interest to us as it lies outside the domain of dependence of
$\overline{\mathcal{S}} $.

\bigskip
As $\overline{\mathcal{S}}$ is a compact set, there is a point in
the interior of $\overline{\mathcal{S}}$ for which $D_i\Omega=0$ and hence $ \Xi = 0$. If
the data is close enough to Minkowski data this critical point is
unique. If one makes the choice $\alpha=0$, then one finds that
\[
\tau_\pm=\pm \frac{2\Omega}{\Xi} \longrightarrow \infty \quad \mbox{ as } \quad \Xi\longrightarrow 0 \quad \mbox{ while } \quad \Omega > 0.
\]
Thus, in order to have a conformal representation of the domain of
dependence of $\overline{\mathcal{S}}$ for which $\tau$ remains
overall finite on $\scri^\pm$ one needs $\alpha\neq 0$.

\bigskip In order to identify points which can be regarded as
representing \emph{timelike infinity}, one needs to investigate the
critical points of $\Theta$, that is, the points for which
$\mbox{d}\Theta=0$. One has that
\begin{eqnarray*}
&& \mbox{d}\Theta = \left( 1 + \alpha \tau + \left( \frac{1}{4}\alpha^2 -\frac{1}{\omega^2}  \right) \tau^2  \right) \mbox{d} \Omega  \nonumber\\
&& \hspace{1.5cm}+ \Omega \left(\alpha + 2\tau \left( \frac{1}{4}\alpha^2-\frac{1}{\omega^2}  \right)  \right) \mbox{d}\tau + \Omega\tau \mbox{d}\alpha + \Omega\tau^2 \left( \frac{1}{2}\alpha \mbox{d}\alpha + \frac{2}{\omega^3}\mbox{d}\omega  \right). \nonumber
\end{eqnarray*}
In particular, we are interested in analysing the conditions
$\mbox{d}\Theta=0$ on $\mathscr{I}^\pm$, with $\Omega\neq 0$. By
construction one finds
\[
\left( 1 + \alpha \tau_\pm + \left( \frac{1}{4}\alpha^2 -\frac{1}{\omega^2}  \right) \tau_\pm^2  \right)  =0.
\]
A necessary condition for the vanishing of $\mbox{d}\Theta$ on $\mathscr{I}^\pm$ for $\Omega\neq 0$ is that
\[
\alpha + 2\tau_\pm \left( \frac{1}{4}\alpha^2-\frac{1}{\omega^2}  \right)=0.
\]
A short calculation shows that the latter is equivalent to
\[
\Xi^2 = D_k \Omega D^k \Omega =0, 
\]
so that $\quad D_k \Omega =0$. Now, if $D_k \Omega=0$ then
$\tau_\pm=-2/ \alpha$. Note that $\tau_\pm>0$ if $\alpha<0$ ---that
is, if $\dot{\Theta}_*<0$. Thus, in order to consider a conformal
representation which includes the point $i^+$, one needs to consider
$\alpha\neq 0$. This condition will be assumed in the sequel. Let
\[
\tau_{i^+} \equiv -2/\alpha.
\]
From the discussion in section \ref{section:Minkowski:reference} it
follows that in particular for the development of Minkowski data one
has that $\tau_{i^+}=2$.

\medskip
Another computation shows that
\[
\Omega\tau_{i^+} \mbox{d}\alpha + \Omega\tau_{i^+}^2 \left( \frac{1}{2}\alpha \mbox{d}\alpha + \frac{2}{\omega^3}\mbox{d}\omega  \right)=0
\]
if $\mbox{d}\Omega=0$. To show this, one uses the fact that if $D_k
\Omega=0$ and $\Omega\neq 0$ then it follows that $D_k(\Xi^2)=0$ and
furthermore, that $1/\omega =\Xi/2\Omega=0$. In view of this, one
defines $i^+ \in \Real \times \mathcal{S}$ as the unique point for
which $\tau =\tau_{i^+}$ and $\mbox{d}\Omega=0$.

\bigskip
To conclude the discussion of the point $i^+$, we look at the
Hessian of $\Theta$ at $i^+$, using the conformal Gaussian coordinates $(\tau, x^\mathcal{A})$. 
Using $D_k \Omega=0$ and $D_k(\Xi^2)=0$ from above, one has 
\[
\nabla_{{\mathcal{A}}} \nabla_{{\mathcal{B}}} \left( \frac{1}{\omega^2}   \right) =  \frac{ \nabla_{{\mathcal{A}}} \nabla_{{\mathcal{B}}} \Xi^2}{4\Omega^2} \quad  \mbox{ at } i^+ . 
\]
Thus we find that 
\begin{eqnarray*}
&&\nabla_{{\mathcal{A}}} \nabla_{{\mathcal{B}}} \Theta \vert_{i^+}= \Omega \nabla_{{\mathcal{A}}} \nabla_{{\mathcal{B}}}\left(\alpha \tau +  \left(\frac{1}{4} \alpha^2 -\frac{1}{\omega^2}\right)\tau^2\right)  \nonumber  \\
&&\phantom{\nabla_{{\mathcal{A}}} \nabla_{{\mathcal{B}}} \Theta \vert_{i^+}}= \Omega\tau \nabla_{{\mathcal{A}}} \nabla_{{\mathcal{B}}} \alpha + \Omega \tau^2\left( \frac{1}{2} \nabla_{{\mathcal{A}}}\alpha \nabla_{{\mathcal{B}}}\alpha + \frac{1}{2} \alpha \nabla_{{\mathcal{A}}} \nabla_{{\mathcal{B}}} \alpha - 
 \frac{\nabla_{{\mathcal{A}}} \nabla_{{\mathcal{B}}}  \Xi^2}{4\Omega^2}  
\right).  \nonumber
\end{eqnarray*}
Consequently 
\begin{equation}
\label{DDTheta_1}
\nabla_{{\mathcal{A}}}\nabla_{{\mathcal{B}}} \Theta 
= \frac{2\Omega}{\alpha^2} \nabla_{{\mathcal{A}}} \alpha \nabla_{{\mathcal{B}}}\alpha
- \frac{ \nabla_{{\mathcal{A}}} \nabla_{{\mathcal{B}}} \Xi^2}{\Omega \alpha^2} 
, \quad \mbox{ on } i^+.
\end{equation}
Similar direct calculations render
\begin{subequations}
\begin{eqnarray}
&& \nabla_{{\mathcal{A}}} \nabla_0 \Theta= \nabla_0 \nabla_{{\mathcal{A}}}\Theta = -\Omega \nabla_{{\mathcal{A}}} \alpha, \quad \mbox{ on } i^+, \label{DDTheta_2}\\
&& \nabla_0 \nabla_0 \Theta = \frac{1}{2} \Omega \alpha^2, \quad \mbox{ on } i^+. \label{DDTheta_3}
\end{eqnarray}
\end{subequations}
Thus, if one chooses the function $\alpha$ such that $\alpha\neq 0$
(in order to have $\tau_{i+}$ finite) and such that the $4\times 4$
matrix with entries given by equations (\ref{DDTheta_1}),
(\ref{DDTheta_2}) and (\ref{DDTheta_3}) has no zero eigenvalues, it
follows that the Hessian of $\Theta$ on $i^+$ is
non-degenerate. Accordingly, the point $i^+$ can be rightfully
regarded as the timelike infinity of the development of the
hyperboloidal data prescribed on $\overline{\mathcal{S}}$.

\section{Evolution equations}
\label{section:evolution:eqns}

The general conformal Einstein field equations introduced in
\cite{Fri95} ---see also \cite{Fri03a,Fri04}--- are a generalisation
of the original conformal equations which allows for the use of Weyl
connections. The use of Weyl connections makes it possible to
consider more general gauges when one is confronted with the need of
deriving a system of propagation equations out of the conformal field
equations. In particular, it makes it possible to make use of the
conformal Gaussian coordinates discussed in the previous sections.

\subsection{Evolution equations in a frame formalism}
As shown in, for example \cite{Fri04}, the general conformal field
equations together with the gauge given by (\ref{cggauge}) imply the
following system of propagation equations:
\begin{eqnarray*}
&&\partial_\tau e\tensor{}{\es}{i} = - \hat{\Gamma}\tensor{i}{p}{0} e\tensor{}{\es}{p} , \nonumber\\
&&\partial_\tau  \hat{\Gamma}\tensor{j}{k}{l}=- \hat{\Gamma}\tensor{q}{k}{l}  \hat{\Gamma}\tensor{j}{q}{0} +  \delta\tensor{}{k}{0}\hat{L}_{jl} - \delta\tensor{}{k}{l}\hat{L}_{j0} -g_{l0}\hat{L}\tensor{j}{k}{} + \Theta d\tensor{0j}{k}{l} , \nonumber\\
&&\partial_\tau \hat{L}_{jl} =- \hat{\Gamma}\tensor{j}{p}{0} \hat{L}_{pl} +  d_k d\tensor{0j}{k}{l}   , \nonumber\\
&&\nabla_k d\tensor{ij}{k}{l} = 0, \nonumber
  \end{eqnarray*}
where $d_{ijkl}=\Theta^{-1} C_{ijkl}$ denotes the components of the
\emph{rescaled Weyl tensor} with respect to the frame $e_{\bar{s}}
$, while the conformal factor $\Theta$ and the 1-form $d_k$ are
given by (\ref{canonical_Theta1})-(\ref{canonical_d}).

\subsection{Evolution equations in a spinor formalism}
In view of future applications, instead of the frame evolution equations discussed in the previous section, we shall use a spinorial version thereof. Using a spin dyad $\{\o, \iota\}$ such that
\[
\tau^{A\Ad} = \sqrt{2} v^{A\Ad} = \o^A \bar{\o}^\Ad +  \i^A \bar{\i}^\Ad,
\]
the evolution equations take the form
 \begin{eqnarray*}
&&\sqrt{2}\partial_\tau e\tensor{}{\es}{C\Cd} = - \hat{\Gamma}\tensor{C\Cd}{E\Ed}{D\Dd} \tau^{D\Dd} e\tensor{}{\es}{E\Ed} , \nonumber\\
&&\sqrt{2}\partial_\tau  \hat{\Gamma}\tensor{D\Dd}{A}{B}  =- (\hat{\Gamma}\tensor{D\Dd}{F}{C}  \hat{\Gamma}\tensor{F\Cd}{A}{B} + \bar{\hat{\Gamma}}\tensor{D\Dd}{F}{C} \hat{\Gamma}\tensor{F \Cd}{A}{B}) \tau^{C\Cd}  \nonumber \\
&& \hspace{5cm}+  \hat{L}_{D\Dd B\Cd} \tau^{A\Cd}  + \Theta \phi\tensor{}{A}{BCD}\tau\tensor{}{C}{\Dd} ,\nonumber \\
&&\sqrt{2}\partial_\tau \hat{L}_{A\Ad C\Cd}  =- (\hat{\Gamma}\tensor{A\Ad}{F}{B}    \hat{L}_{F\Bd C\Cd} + \bar{\hat{\Gamma}}\tensor{A\Ad}{\Fd}{\Bd}    \hat{L}_{B\Fd C\Cd})\tau^{B\Bd} \nonumber \\
&& \hspace{5cm}- d^{E\Ed} (\phi_{EABC} \epsilon_{\Ed\Cd}\tau\tensor{}{B}{\Ad} + \bar{\phi}_{\Ed\Ad\Bd\Cd} \epsilon_{EC}\tau\tensor{A}{\Bd}{} )  ,\nonumber\\
&&\sqrt{2}\partial_\tau \phi_{ABCD} =\tau^{F\Fd} \nabla_{\Fd(D}\phi_{ABC)F} - \tau_{\Fd(D} \nabla^{F\Fd}\phi_{ABC)F}, \nonumber
  \end{eqnarray*}
where one has the correspondences (via the Infeld-van der Waerden symbols)
\begin{eqnarray*}
&& e\tensor{}{\es}{i} \mapsto e\tensor{}{\es}{C\Cd}, \nonumber \\
&& \hat{\Gamma}\tensor{j}{k}{l} \mapsto \dnupdn{\hat{\Gamma}}{BB'}{AA'}{CC'}=\dnupdn{\hat{\Gamma}}{BB'}{A}{C} \dnup{\epsilon}{C'}{A'} + \dnupdn{\bar{\hat{\Gamma}}}{BB'}{A'}{C'} \dnup{\epsilon}{C}{A}, \nonumber \\
&& \hat{L}_{ij} \mapsto \hat{L}_{AA'BB'}, \nonumber \\
&& d_{ijkl} \mapsto d_{AA'BB'CC'DD'}= \phi_{ABCD} \epsilon_{A'B'}\epsilon_{C'D'} + \bar{\phi}_{A'B'C'D'}\epsilon_{AB}\epsilon_{CD}, \nonumber \\
&& d_j \mapsto d_{AA'},\nonumber
\end{eqnarray*}
and the factors of $\sqrt{2}$ arise from the normalisation $\tau_{AA'}
\tau^{AA'}=2$.

\subsection{Evolution equations in space-spinor formalism}

Next, one introduces a \emph{space-spinor formalism} by using the
spinor $\tau\tensor{A}{\Ad}{}$ to eliminate all primed indices and
then one splits the equations into symmetric and skew parts.

\bigskip
A space spinor $\Theta_{ABCD}=\Theta_{AB(CD)}$ is
introduced and then decomposed such that
\[
\Theta_{ABCD} \equiv \hat{L}_{C\Cd A\Ad} \updn{\tau}{A'}{B} \updn{\tau}{C'}{D}=\Theta_{(AB)(CD)} + \frac{1}{2} \epsilon_{AB} \dnupdn{\Theta}{G}{G}{(CD)}.
\]
For the spin coefficients $\hat{\Gamma}_{AA'BC}$, one observes that $\hat{\Gamma}_{AA'BC}=\Gamma_{AA'BC}+\epsilon_{AB}f_{CA'}$ and defines
\[
\Gamma_{ABCD} \equiv \dnup{\tau}{B}{B'} \Gamma_{AB'CD},
\]
which in turn, will be decomposed as 
\[
\Gamma_{ABCD} = \frac{1}{\sqrt{2}}\left(\xi_{ABCD} - \chi_{(AB)CD} \right) -\frac{1}{2}\epsilon_{AB} f_{CD}.
\]
The spinors in the latter equation possess the following symmetries
\[
\Gamma_{ABCD}=\Gamma_{AB(CD)}, \quad \chi_{ABCD}=\chi_{AB(CD)}, \quad \xi_{ABCD}=\xi_{(AB)(CD)}.
\]
The term $\xi_{ABCD}$ is related to the intrinsic connection of the
leaves of the foliation defined by $\tau_{AA'}$; the term
$\chi_{(AB)CD}$ to the second fundamental form of the leaves; and
$f_{AB}$ to the acceleration of the foliation. The spinors
$\xi_{ABCD}$, $\chi_{(AB)CD}$ and $f_{AB}$ are calculated from
$\Gamma_{ABCD}$ via the relations
\begin{eqnarray*}
&& \xi_{ABCD} = \frac{1}{\sqrt{2}} \left( \Gamma_{ABCD} + \dnup{\tau}{B}{B'} \dnup{\tau}{C}{C'} \dnup{\tau}{D}{D'} \overline{\Gamma}_{AB'C'D'}  \right), \nonumber\\
&& \chi_{ABCD} = \frac{1}{\sqrt{2}} \left(\dnup{\tau}{B}{B'} \dnup{\tau}{C}{C'} \dnup{\tau}{D}{D'} \overline{\Gamma}_{AB'C'D'}-\Gamma_{ABCD}  \right), \nonumber\\ 
&& f_{AB}=- \tau^{CC'} \Gamma_{CC'AB}. \nonumber
\end{eqnarray*}
The frame fields $e^\es_{AA'}$ are decomposed using
\begin{eqnarray*}
&& e^0_{AA'}=\frac{1}{\sqrt{2}}\tau_{AA'} - \updn{\tau}{B}{A'}e^0_{AB}, \nonumber\\
&& e^\er_{AA'} = -\updn{\tau}{B}{A'}e^\er_{AB},\nonumber
\end{eqnarray*}
with
\[
e^\es_{AB}\equiv \dnup{\tau}{(A}{B'} e^\es_{B)B'}.
\]
The fields $e^\es_{AB}$ are associated to spatial vectors, and hence,
they satisfy the reality conditions
\begin{equation}
e^\es_{AB} = -\dnup{\tau}{A}{A'}
\dnup{\tau}{B}{B'}\overline{e}^\es_{A'B'}. \label{reality:condition}
\end{equation}

\bigskip
Using the gauge given by (\ref{cggauge}) it can be shown that the
extended conformal field equations given in \cite{Fri98a} imply the
following evolution equations for the unknowns $e^\es_{AB}$,
$\xi_{ABCD}$, $f_{AB}$, $\chi_{(AB)CD}$, $\Theta_{(AB)CD}$,
$\Theta_{G\phantom{G}CD}^{\phantom{G}G}$:
\begin{subequations}
\begin{eqnarray}
&&\partial_\tau e^0_{AB}=-\dnup{\chi}{(AB)}{EF}e^{0}_{EF}-f_{AB}, \label{p1} \\
&&\partial_\tau e^\er_{AB}=-\dnup{\chi}{(AB)}{EF}e^\er_{EF}, \label{p2}\\
&&\partial_\tau \xi_{ABCD}=-\dnup{\chi}{(AB)}{EF}\xi_{EFCD}+\frac{1}{\sqrt{2}}(\epsilon_{AC}\chi_{(BD)EF}+\epsilon_{BD}\chi_{(AC)EF})f^{EF} \nonumber\\
&&\hspace{2cm} -\sqrt{2}\dnup{\chi}{(AB)(C}{E}f_{D)E}-\frac{1}{2}(\epsilon_{AC}\dnupdn{\Theta}{F}{F}{BD}+\epsilon_{BD}\dnupdn{\Theta}{F}{F}{AC})-\mbox{i}\Theta\mu_{ABCD}, \label{p3} \\
&&\partial_\tau f_{AB}=-\dnup{\chi}{(AB)}{EF}f_{EF}+\frac{1}{\sqrt{2}}\dnupdn{\Theta}{F}{F}{AB}, \label{p4} \\
&&\partial_\tau \chi_{(AB)CD}=-\dnup{\chi}{(AB)}{EF}\chi_{EFCD}-\Theta_{(CD)AB}+\Theta\eta_{ABCD}, \label{p5} \\
&&\partial_\tau\Theta_{(AB)CD}=-\dnup{\chi}{(CD)}{EF}\Theta_{(AB)EF}-\partial_\tau\Theta\eta_{ABCD}+\mbox{i}\sqrt{2}\updn{d}{E}{(A}\mu_{B)CDE}, \label{p6} \\
&&\partial_\tau \dnupdn{\Theta}{G}{G}{AB}=-\dnup{\chi}{(AB)}{EF}\dnupdn{\Theta}{G}{G}{EF}+\sqrt{2}d^{EF}\eta_{ABEF}, \label{p7}
\end{eqnarray}
\end{subequations}
where
\begin{eqnarray*}
&&\eta_{ABCD}=\frac{1}{2}(\phi_{ABCD}+\dnup{\tau}{A}{A'}\dnup{\tau}{B}{B'}\dnup{\tau}{C}{C'}\dnup{\tau}{D}{D'} \overline{\phi}_{A'B'C'D'}), \nonumber \\
&&\mu_{ABCD}=-\frac{\mbox{i}}{2}(\phi_{ABCD}-\dnup{\tau}{A}{A'}\dnup{\tau}{B}{B'}\dnup{\tau}{C}{C'}\dnup{\tau}{D}{D'}\overline{\phi}_{A'B'C'D'}), \nonumber
\end{eqnarray*}
denote, respectively, the electric and magnetic parts of of
$\phi_{ABCD}$.

\bigskip
The evolution equations for the spinor $\phi_{ABCD}$ are derived from
the Bianchi equations. Depending on the need, several
alternative systems can be deduced. Here, we will consider the one
which was called the \emph{standard system} in references
\cite{Fri95,Fri98a}. Let
\[
\phi_{ABCD}= \phi_0 \epsilon^0_{ABCD} + \phi_1 \epsilon^1_{ABCD} + \phi_2 \epsilon^2_{ABCD} + \phi_3 \epsilon^3_{ABCD} = \phi_4 \epsilon^4_{ABCD},
\]
where
\begin{eqnarray*}
&&\phi_i \equiv \phi_{(ABCD)_i}, \quad i=0,\ldots,4, \nonumber \\
&&\epsilon^0_{ABCD}\equiv o_{(A}o_B o_C o_{D)}, \quad \epsilon^1_{ABCD} \equiv \iota_{(A}o_B o_C o_{D)}, \quad \epsilon^2_{ABCD} \equiv \iota_{(A} \iota_B o_C o_{D)}, \nonumber \\
&& \epsilon^3_{ABCD} \equiv \iota_{(A} \iota_B \iota_C o_{D)}, \quad \epsilon^4_{ABCD} \equiv \iota_{(A} \iota_B \iota_C \iota_{D)}. \nonumber
\end{eqnarray*}

In the previous expressions the subindex ${}_{(ABCD)_i}$ indicates that
after the symmetrisation $i$ indices are set to zero. One has the
following
\emph{Bianchi propagation equations}:
\begin{subequations}
\begin{eqnarray}
&&(\sqrt{2}-2e^0_{01})\partial_\tau\phi_0+2e^0_{00}\partial_\tau\phi_1-2e^\er_{01}\partial_{\er}\phi_0+2e^\er_{00}\partial_\er\phi_1 \nonumber \\
&&\hspace{1cm}= (2\Gamma_{0011}-8\Gamma_{1010})\phi_0+(4\Gamma_{0001}+8\Gamma_{1000})\phi_1-6\Gamma_{0000}\phi_2, \label{b0}\\
&&\sqrt{2}\partial_\tau\phi_1-e^0_{11}\partial_\tau\phi_0+e^0_{00}\partial_\tau\phi_2-e^\er_{11}\partial_{\er}\phi_0+e^\er_{00}\partial_{\er}\phi_2\nonumber\\
&&\hspace{1cm}=-(4\Gamma_{1110}+f_{11})\phi_0+(2\Gamma_{0011}+4\Gamma_{1100}-2f_{01})\phi_1
+3f_{00}\phi_2-2\Gamma_{0000}\phi_3, \label{b1} \\
&&\sqrt{2}\partial_\tau\phi_2-e^0_{11}\partial_\tau\phi_1+e^0_{00}\partial_\tau\phi_3-e^\er_{11}\partial_{\er}\phi_1+e^\er_{00}\partial_{\er}\phi_3\nonumber \\
&&\hspace{1cm}=-\Gamma_{1111}\phi_0-2(\Gamma_{1101}+f_{11})\phi_1+3(\Gamma_{0011}+\Gamma_{1100})\phi_2 \nonumber \\
&&\hspace{2cm}-2(\Gamma_{0001}-f_{00})\phi_3-\Gamma_{0000}\phi_4, \label{b2}\\
&&\sqrt{2}\partial_\tau\phi_3-e^0_{11}\partial_\tau\phi_2+e^0_{00}\partial_\tau\phi_4-e^\er_{11}\partial_{\er}\phi_2+e^\er_{00}\partial_{\er}\phi_4\nonumber \\
&&\hspace{1cm}=-2\Gamma_{1111}\phi_1
-3f_{11}\phi_2+(2\Gamma_{1100}+4\Gamma_{0011}+2f_{01})\phi_3-(4\Gamma_{0001}-f_{00})\phi_4,
\label{b3}\\
&&(\sqrt{2}+2e^0_{01})\partial_\tau\phi_4-2e^0_{11}\partial_\tau\phi_3+2e^\er_{01}\partial_\er\phi_4-2e^\er_{11}\partial_\er\phi_3 \nonumber \\
&&\hspace{1cm}=-6\Gamma_{1111}\phi_2+(4\Gamma_{1110}+8\Gamma_{0111})\phi_3
+(2\Gamma_{1100}-8\Gamma_{0101})\phi_4. \label{b4}
\end{eqnarray}
\end{subequations}

\subsection{Evolution equations for the Jacobi field}
To the above propagation equations we will have to append an evolution
equation for the Jacobi field. The conformal Jacobi field $\eta^\mu$
has a spinorial counterpart $\eta_{AA'}$ which can be split as
\[
\eta_{AA'} = \frac{1}{2}\eta \tau_{AA'} -\updn{\tau}{B}{A'} \eta_{AB},
\]
with
\[
\eta \equiv \eta_{AA'} \tau^{AA'}, \quad \eta_{AB}\equiv \dnup{\tau}{(A}{B'} \eta_{B)B'}.
\]
Conjugate points in the congruence of conformal geodesics arise if
$\eta_{AB}=0$. The components $\eta$, $\eta_{AB}$ satisfy the
propagation equations
\begin{subequations}
\begin{eqnarray}
&& \sqrt{2} \partial_\tau \eta= f_{AB} \eta^{AB}, \label{j1}\\
&& \sqrt{2} \partial_\tau \eta_{AB} = \chi_{CD(AB)} \eta^{CD}. \label{j2}
\end{eqnarray}
\end{subequations}

\subsection{Structural properties of the evolution equations}
\label{section:structural:properties}
We discuss now some general structural properties of the equations
(\ref{p1})-(\ref{p7}), (\ref{b1})-(\ref{b4}), (\ref{j1})-(\ref{j2})
which will be used systematically in the sequel. Introduce the
notation
\[
\upsilon \equiv \left(e^\es_{AB}, \Gamma_{ABCD}, \Theta_{ABCD}\right), \quad \phi\equiv \left(\phi_0,\phi_1,\phi_2,\phi_3,\phi_4\right),
\]
where it is understood that $\upsilon$ contains only the independent
components of the respective spinor ---which are obtained by writing
linear combinations of irreducible spinors, as discussed in the
previous section. The unknown vector $\upsilon$ has 45 independent
complex components, while $\phi$ has 5 complex components. In terms of
$\upsilon$ and $\phi$, the propagation equations (\ref{p1})-(\ref{p7})
and (\ref{j1})-(\ref{j2}) can be written as:
\begin{equation}
\label{upsilon:propagation}
\partial_\tau \upsilon = K\upsilon + Q(\upsilon,\upsilon)+ L\phi,
\end{equation}
where $K$ and $Q$ denote , respectively, a linear constant
matrix-valued function and a bilinear vector-valued function both with
constant entries and $L$ is a linear matrix-valued function with
coefficients depending on the coordinates.  Similarly the system
(\ref{b1})-(\ref{b4}) can be written as
\begin{equation}
\sqrt{2}E \partial_\tau \phi + A^{AB} e^\es_{AB}\partial_\es \phi =B(\Gamma_{ABCD})\phi, \label{bianchi:propagation}
\end{equation}
where $E$ denotes the $5\times 5$ identity matrix and $A^{AB}e^\es_{AB}$,
$\es=0,\ldots,3$, are $5\times 5$ matrices depending on the
coordinates, while $B(\Gamma_{ABCD})$ denotes a constant matrix-valued
linear function of the connection coefficients $\Gamma_{ABCD}$. For
later reference it is noted that
\[
\sqrt{2}E + A^{AB}e^0_{AB}=
\left(
\begin{array}{ccccc}
\sqrt{2}-2e^0_{01} & 2e^0_{00} & 0 & 0 & 0 \\
-e^0_{11} & \sqrt{2} & e^0_{00} & 0 & 0 \\
0 & -e^0_{11} & \sqrt{2} & e^0_{00} & 0 \\
0 & 0 & -e^0_{11} & \sqrt{2} & e^0_{00} \\
0 & 0 & 0 & -2e^0_{11} & \sqrt{2} + 2e^0_{01}
\end{array}
\right),
\]
and that
\[
A^{AB}e^\er_{AB}=
\left(
\begin{array}{ccccc}
-2e^\er_{01} & 2e^\er_{00} & 0 & 0 & 0 \\
-e^\er_{11} & \sqrt{2} & e^\er_{00} & 0 & 0 \\
0 & -e^\er_{11} & \sqrt{2} & e^\er_{00} & 0 \\
0 & 0 & -e^\er_{11} & \sqrt{2} & e^\er_{00} \\
0 & 0 & 0 & -2e^\er_{11} & 2e^\er_{01}
\end{array}
\right),
\]
with $\er=1,2,3$. From the reality condition (\ref{reality:condition})
one has that
\begin{eqnarray*}
&& e^0_{00}=-\overline{e}^0_{11}, \quad e^\er_{00}=-\overline{e}^\er_{11}, \nonumber \\
&& e^0_{01}=\overline{e}^0_{01}, \quad e^\er_{01}=\overline{e}^\er_{01},\nonumber 
\end{eqnarray*}
so that in particular $e^0_{01}$ and $e^\er_{01}$ are real. Strictly
speaking, a discussion of the symmetric hyperbolicity of the system
(\ref{upsilon:propagation})-(\ref{bianchi:propagation}) should be
carried out using real unknowns. In order to ease the presentation in
the sequel, we write
\[
e^\es_{00}=a^\es +\mbox{i}b^\es, \quad e^\es_{01}=c^\es,
\]
where $a^\es$, $b^\es$ and $c^\es$ denote the components of real
vectors.  Thus, making use of the splitting $\phi_j= \mbox{Re}(\phi_j)
+\mbox{i} \,\mbox{Im}(\phi_j)$ and by multiplying the equations
(\ref{b0})-(\ref{b4}) by suitable numeric constants one finds that the
$5\times 5$ matrix $\sqrt{2}E + A^{AB}e^0_{AB}$ implies the $10\times
10$ matrix:
\[
\tilde{A}^0(a^0,b^0,c^0)=\left(
\begin{array}{cccccccccc}
\displaystyle \frac{1}{\sqrt{2}}-c^0 & a^0 & 0 & 0 & 0 & 0 & -b^0 & 0 & 0 & 0 \\
a^0 & \sqrt{2} & a^0 & 0 & 0 & b^0 & 0 & -b^0 & 0 & 0 \\
0 & a^0 & \sqrt{2} & a^0 & 0 & 0 & b^0 & 0 & -b^0 & 0 \\
0 & 0 & a^0 & \sqrt{2} & a^0 & 0 & 0 & b^0 & 0 & -b^0 \\
0 & 0 & 0 & a^0 & \displaystyle \frac{1}{\sqrt{2}}+c^0 & 0 & 0 & 0 & b^0 & 0  \\
0 & b^0 & 0 & 0 & 0 & \displaystyle \frac{1}{\sqrt{2}}-c^0 & a^0 & 0 & 0& 0 \\
-b^0 & 0 & b^0 & 0 & 0 & a^0 & \sqrt{2} & a^0 & 0 & 0 \\
0 & -b^0 & 0 & b^0 & 0 & 0 & a^0 & \sqrt{2} & a^0 & 0 \\
0 & 0 & -b^0 & 0 & b^0 & 0 & 0 & a^0 & \sqrt{2} & a^0 \\
0 & 0 & 0 & -b^0 & 0 & 0 & 0 & 0 & a^0 & \displaystyle \frac{1}{\sqrt{2}}+c^0
\end{array}
\right).
\]
In particular note that if one sets $a^0=b^0=c^0=0$, then one gets
\begin{equation}
\label{reference:submatrix}
\tilde{A}^0(0,0,0)=\mbox{diag}\left(\frac{1}{\sqrt{2}}, \sqrt{2}, \sqrt{2},\sqrt{2},\frac{1}{\sqrt{2}}, \frac{1}{\sqrt{2}}, \sqrt{2}, \sqrt{2},\sqrt{2},\frac{1}{\sqrt{2}}  \right).
\end{equation}
Similarly, from the $5\times5$ matrix $A^{AB}e^\er_{AB}$ one deduces
the (real) $10 \times 10$ symmetric matrix:
\[
\tilde{A}^\er(a^\er,b^\er,c^\er)=\left(
\begin{array}{cccccccccc}
-c^\er & a^\er & 0 & 0 & 0 & 0 & -b^\er & 0 & 0 & 0 \\
a^\er & \sqrt{2} & a^\er & 0 & 0 & b^\er & 0 & -b^\er & 0 & 0 \\
0 & a^\er & \sqrt{2} & a^\er & 0 & 0 & b^\er & 0 & -b^\er & 0 \\
0 & 0 & a^\er & \sqrt{2} & a^\er & 0 & 0 & b^\er & 0 & -b^\er \\
0 & 0 & 0 & a^\er & c^\er & 0 & 0 & 0 & b^\er & 0  \\
0 & b^\er & 0 & 0 & 0 & -c^\er & a^\er & 0 & 0& 0 \\
-b^\er & 0 & b^\er & 0 & 0 & a^\er & \sqrt{2} & a^\er & 0 & 0 \\
0 & -b^\er & 0 & b^\er & 0 & 0 & a^\er & \sqrt{2} & a^\er & 0 \\
0 & 0 & -b^\er & 0 & b^\er & 0 & 0 & a^\er & \sqrt{2} & a^\er \\
0 & 0 & 0 & -b^\er & 0 & 0 & 0 & 0 & a^\er & c^\er
\end{array}
\right).
\]
For each $\es=0,1,2,3$, the matrices $\tilde{A}^\es(z) $ have entries
which are polynomials of at most degree one in $z=(a^\es,b^\es,c^\es)
$. We can rewrite them in the following form
\[ 
\tilde{A}^\es(z) = \tilde{A}^\es(0) + \check{\tilde{A}}^\es(z)
\]
where $\tilde{A}^\es(0) \equiv \tilde{A}^\es(0,0,0)$ and
$\check{\tilde{A}}^\es(x+y)= \check{\tilde{A}}^\es(x)+\check{\tilde{A}}^\es(y)$.

\section{The conformal de Sitter and  Minkowski space-times as solutions to the conformal field equations}
\label{reference:st:as:solns}

In the conformal geodesic gauge given by (\ref{cggauge}) both the de
Sitter and the Minkowski space-times are conformally rescaled to the
unphysical space-time $ (\mathcal{M},g)$ where $g= \Theta^2_E g_E $
---see equation (\ref{g_E}). The connection and curvature spinor
components form the variables $\upsilon$ and $\phi$ that satisfy the
equations (\ref{upsilon:propagation}) and (\ref{bianchi:propagation})
---respectively (\ref{p1})-(\ref{p7}) and (\ref{b0})-(\ref{b4})--- can
be directly calculated from the components for the Einstein cylinder.
More precisely, a straightforward calculation using the results of
section \ref{asimple:model:solns} renders
\begin{subequations}
\begin{eqnarray}
&& e^0_{AB} = 0, \label{Ecosmos1}\\
&& e^\er_{AB} = \frac{4}{\tau^2 + 4} \sigma^{\er}_{AB},\\
&& f_{AB}= 0 ,\\
&& \xi_{ABCD} = \frac{-2\mbox{i}}{\tau^2 + 4} h_{ABCD}, \\
&& \chi_{(AB)CD} = \frac{2\tau}{\tau^2 + 4} h_{ABCD}, \\
&& \Theta_{ABCD}=  \frac{-2}{\tau^2 + 4} h_{ABCD} , \\
&& \phi_{ABCD}= 0, \label{Ecosmos7}
\end{eqnarray}
\end{subequations}
where $\sigma^{\er}_{AB}\equiv
\sigma^{\er}_{\Ad (A}\dnup{\tau}{B)}{A'} $ are the spatial Infeld-van
der Waerden symbols and
\[
h_{ABCD}\equiv - \epsilon_{A(C}\epsilon_{D)B}.
\] 
The solutions to the conformal Jacobi equations (\ref{j1}) and
(\ref{j2}) for the reference solution are given by
\begin{subequations}
\begin{eqnarray}
&& \eta =0, \label{Ecosmos8}\\
&& \eta^\er_{AB}= \left(1+ \frac{\tau^2}{4} \right) \sigma^\er_{AB}. \label{Ecosmos9}
\end{eqnarray}
\end{subequations}

\bigskip It is important to note that expressions
(\ref{Ecosmos1})-(\ref{Ecosmos7}) and
(\ref{Ecosmos8})-(\ref{Ecosmos9}) are valid for both the conformal de
Sitter and the conformal Minkowski space-times. What distinguishes
these two conformal space-times is the use of the appropriate
conformal factor $\Theta_D$, resp. $\Theta_M$ ---cfr. section
\ref{asimple:model:solns}. It should also be observed that the
expressions (\ref{Ecosmos1})-(\ref{Ecosmos7}) and
(\ref{Ecosmos8})-(\ref{Ecosmos9}) are analytic functions of $\tau$ for
$\tau\in \Real$. Important in the sequel will be that $\eta^\er_{AB}$
are non-vanishing for $\tau\in(-\infty,\infty)$.

\section{The existence and stability results}
\label{section:existence:results}
Having written the evolution equations in the conformal geodesic
gauge, we proceed now to analyse the existence of solutions close to
the explicit reference solutions of section
\ref{reference:st:as:solns}. Following the original discussion in
\cite{Fri86b}, the desired existence and stability results are
obtained making use of slight modifications of very general theorems
by Kato on the properties of symmetric hyperbolic systems
\cite{Kat70,Kat73,Kat75}.

\subsection{Some further structural properties of the propagation system}
Let
$u=(\mbox{Re}(\upsilon),\mbox{Im}(\upsilon),\mbox{Re}(\phi),\mbox{Im}(\phi))$
with $\upsilon$ and $\phi$ as in section
\ref{section:structural:properties}. The unknown $u$ takes values in
$\Real^N$ for some $N\in \Natural$. The evolution equations
(\ref{p1})-(\ref{p7}) and (\ref{b0})-(\ref{b4}) ---or their matricial
counterparts (\ref{upsilon:propagation}) and
(\ref{bianchi:propagation})--- render a system of quasilinear partial
differential equations for $u$ which has the form
\begin{equation}
A^0(u) \cdot \partial_\tau u+ \sum_{\er=1}^3 A^\er(u)\cdot c_\er(u) +B(\tau,x^\mathcal{A},u)\cdot u=0,
\label{original:system}
\end{equation}
with $c_\er(u)$ denoting the vector fields
(\ref{v_field_c1})-(\ref{v_field_c3}) acting on the unknown $u$.
Furthermore,
\[
A^0(u)=
\left(
\begin{array}{cc}
E & 0 \\
0 & \tilde{A}^0(u)
\end{array}
\right),
\quad
A^\er(u)=
\left(
\begin{array}{cc}
0 & 0 \\
0 & \tilde{A}^\er(u)
\end{array}
\right),
\]
with $E$, $0$ denoting, respectively, identity and zero matrices of
the suitable dimensions and $\tilde{A}^0(u)$ and $\tilde{A}^\er(u)$ as
given in section \ref{section:structural:properties}. In particular given any $z\in \Real^N $,
the matrix valued functions $A^\es(z)$, $\es=0,1,2,3$ have entries which
are polynomial in $z$. These polynomials are at most of
degree one and have constant coefficients. The matrices are symmetric
${}^t(A^\es(z))=A^\es(z)$, $z\in\Real^N$. The matrix valued function
$B=B(\tau,x^\mathcal{A},z)$ with $(\tau,x^\mathcal{A},z)\in
\Real\times \mathcal{S} \times \Real^N$, has entries which are
polynomials in $z$ with coefficients which are analytic functions on
$\Real \times \mathcal{S}$. Note that as $\mathcal{S}$ is
diffeomorphic to $\Sphere^3$, then $B$ can be regarded as a matrix
valued function with domain $\Real \times \Sphere^3 \times \Real^N$
---this point of view will be adopted in what follows. These
polynomials are at most of degree one. Following the decomposition in section 
\ref{section:structural:properties}, one can write 
\begin{eqnarray*}
&& A^\es(z) = A^\es(0) + \check{A}^\es(z), \nonumber \\
&& B(\tau,x^\mathcal{A},z)= B(\tau,x^\mathcal{A},0) + \check{B}(\tau, x^\mathcal{A},z), \nonumber
\end{eqnarray*}
with
\begin{eqnarray*}
&& \check{A}^\es(y+z)= \check{A}^\es(y)+\check{A}^\es(z), \nonumber \\
&& \check{B}(\tau, x^\mathcal{A},y+z)=\check{B}(\tau, x^\mathcal{A},y)+\check{B}(\tau, x^\mathcal{A},z).\nonumber
\end{eqnarray*}

\bigskip
Let $u'$ denote the
explicit reference solution given by (\ref{Ecosmos1})-(\ref{Ecosmos7}). Set
\begin{equation}
\label{solution:Ansatz}
u=u'+w.
\end{equation}
This is in essence a definition for the new unknown $w$ ---which gives
the perturbation from the reference solutions. Substitution of the
Ansatz (\ref{solution:Ansatz}) into the system (\ref{original:system})
yields
\[
A^0(u'+w)\cdot \partial_\tau w + \sum_{\er=1}^3 A^\er(u'+w) \cdot c_\er(w) + B(\tau,x^\mathcal{A}, u'+w)\cdot w + \check{A^0}(w)\cdot \partial_\tau u'=0,
\]
where it has been used that
\[
A^0(u')\cdot \partial_\tau + \sum_{\er=1}^3 A^\er(u') \cdot c_\er(u') +B(\tau,x^\mathcal{A},u')\cdot u =0,
\]
and that $c_\er(u')=0$ ---the reference solutions have no spatial
dependence in the gauge being used. Thus, $w$ satisfies
\begin{equation}
\label{sym:hyp:system0}
A^0(u'+w)\cdot \partial_\tau w + \sum_{\er=1}^3 A^\er(u'+w) \cdot c_\er(w) 
+ \widehat{{\mathcal{B}}}(\tau,x^\mathcal{A}, w)\cdot w=0,
\end{equation}
with $ \widehat{{\mathcal{B}}}(\tau,x^\mathcal{A}, z)$ again a matrix valued
function with entries which are polynomials of at most degree one and 
coefficients which are analytic functions on $\Real \times \Sphere^3$,
such that
\[
 \widehat{{\mathcal{B}}}(\tau,x^\mathcal{A}, w)\cdot w =  B(\tau,x^\mathcal{A}, u'+w)\cdot w 
 + \check{A^0}(w)\cdot \partial_\tau u' + \check{B}(\tau,x^\mathcal{A}, w)\cdot u' .
\]
We define new matrices ${\mathcal{A}}^0(w) = A^0(u'+w)$ 
and ${\mathcal{A}}^\er(w) = A^\er(u'+w)$ so that
\begin{equation}
\label{sym:hyp:system}
{\mathcal{A}}^0(w)\cdot \partial_\tau w + \sum_{\er=1}^3 {\mathcal{A}}^\er(w) \cdot c_\er(w) 
+ \widehat{{\mathcal{B}}}(\tau,x^\mathcal{A}, w)\cdot w=0,
\end{equation}
For later use it is noted that the matrix $A^0(u')= {\mathcal{A}}^0(0)$ is diagonal with
constant entries. More precisely, it has the form
\[
{\mathcal{A}}^0(0)=
\left(
\begin{array}{cc}
E & 0 \\
0 & \tilde{A}^0(0)
\end{array}
\right),
\]
with $E$ an identity matrix of the appropriate dimensions and
$\tilde{A}^0(0)$ as given by equation
(\ref{reference:submatrix}). In particular, all the entries of $A^0(u')$
are bigger or equal to $1/\sqrt{2}$.

\subsection{The Kato existence and stability result}
Let $D$ and $\mbox{d}\mu$ denote, respectively, the
Levi-Civita covariant derivative and the volume element associated
with the standard metric on $\Sphere^3$ and $D_\er$ denotes the covariant
derivative in the direction of $c_\er$. On the space
$C^\infty(\Sphere^3,\Real^N)$ of smooth $\Real^N$-valued functions on
$\Sphere^3$ define for $m\in \Natural$ the Sobolev-like norm
\begin{equation}
\label{norm}
\parallel w \parallel_m=\left( \sum_{k=0}^m \int_{\Sphere^3} |D^k w|^2 \mbox{d}\mu \right)^{1/2}.
\end{equation}
The notation
\[
|D^0 w|^2 =|w|^2, \quad |D^k w|^2 = \sum^3_{\er_1,\ldots, \er_k=1} |D_{\er_1}\cdots D_{\er_k}w|^2,
\]
is used, with $|\cdot|$ the standard Euclidean norm on $\Real^N$.
Given $m\in \Natural$ let
$H^m(\Sphere^3,\Real^N)$ be the Hilbert space obtained as the
completion of the space $C^\infty(\Sphere^3,\Real^N)$ in the norm
(\ref{norm}). The unknown $w=w(\tau,x)$ will be regarded as a function of
$\tau$ which takes values in $H^m(\Sphere^3,\Real^N)$. For
$\delta\in\Real$, $m\in\Natural$ with $0<\delta <1/\sqrt{2}$, $m\geq
2$, set
\[
D_\delta^m=\left\{w\in H^m(\Sphere^3,\Real^N) \; |\; (z,{\mathcal{A}}^0(w)z)>\delta(z,z), \forall z\in\Real^N    \right\},
\]
where $(\cdot,\cdot)$ is the standard scalar product on $\Real^N$. Important
for our purposes is that it contains a neighbourhood of the origin of
$H^m(\Sphere^3,\Real^N)$ as the entries of ${\mathcal{A}}^0(0)$ are bounded from below by $1/\sqrt{2}$. 

\bigskip The original existence and stability results by Kato for
symmetric hyperbolic systems of the form (\ref{sym:hyp:system}) which
can be found in \cite{Kat75} ---see also \cite{Kat70,Kat73}--- have
been worked out for the case when the frame fields $c_\er$ commute. In
the case under discussion where the underlying leaves of the foliation
of the space-time have the topology of $\Sphere^3$, the frame fields
$c_\er$ do not commute. As discussed in \cite{Fri86b}, Kato's result
can be modified to handle frame fields whose commutators are those of
$O(3)$. For completeness we quote the result given in \cite{Fri86b}.

\begin{theorem}
\label{thm:mod:Kato}
Suppose $m\geq 4$, $D$ is a bounded open subset of $H^m(\Sphere^3,\Real^N)$ with $D\subset D^m_\delta$. If $w_0\in D$ is given as initial condition for the system (\ref{sym:hyp:system}), then:

\begin{itemize}
\item[(i)] There exists a $T>0$ and a unique solution $w(\tau)$ of equation (\ref{sym:hyp:system}) defined on $[0,T]$ with $w(0)=w_0$ and
\[
w\in C(0,T;D)\cap C^1(0,T; H^{m-1}(\Sphere^3,\Real^N)).
\]

\item[(ii)] There is an $\varepsilon>0$ such that one value for $T$ can be chosen common to all initial conditions in the open ball $B_\varepsilon(w_0)$ with centre $w_0$ and radius $\varepsilon$, and such that $B_\varepsilon(w_0)\subset D$.

\item[(iii)] If the solution $w(\tau)$ in (i) exists on $[0,T_0]$ for some $T_0>0$, then the solutions to all initial conditions in $B_\varepsilon(w_0)$ exist on $[0,T_0]$ if $\varepsilon>0$ is sufficiently small.

\item[(iv)] If $\varepsilon$ and $T$ are chosen as in (ii) and $w_0^n\in B_\varepsilon(w_0)$ with $\parallel w_0^n -w_0\parallel_m\rightarrow 0$ as $n\rightarrow \infty$, then for the solutions $w^n(\tau)$ with $w^n(0)=w_0^n$ it holds that $\parallel w^n(\tau)-w(\tau)\parallel_m\rightarrow 0$ uniformly in $\tau\in[0,T]$ as $n\rightarrow \infty$.
\end{itemize}

\end{theorem}

\textbf{Remark 1.} The point (i) in the theorem establishes the local
existence of solutions to the equation (\ref{sym:hyp:system}) for
sufficiently small data (but not exclusively). By (ii) there is a
non-vanishing existence time common to all solutions arising from data
in a suitably small neighbourhood of the $0$-data ---the reference
solution. In particular, by (iii), if a reference solution is known to
have a certain existence time $T_0$, then all solutions arising from
data sufficiently close to the data of the reference solution have the
same existence time. Finally point (iv) states that data close to a
certain reference data gives rise to developments which are also close
to the reference solution --- i.e. stability.

\medskip
\textbf{Remark 2.} A direct computation shows that on $[0,T]$ the solution to
(\ref{sym:hyp:system}) is of class $H^m((0,T)\times\Sphere^3) \subset
C^{m-2}([0,T]\times \Sphere^3)$. The convergence stated in (iv) is uniform
on $[0,T]\times \Sphere^3$.

\subsection{Application to de Sitter-like space-times}
\subsubsection{Standard Cauchy data for de Sitter-like space-times}
We start by considering the case where on a standard (spacelike)
Cauchy hypersurface $\mathcal{S}$ one is given de Sitter-like initial
data. The initial data consists of the values of the spinorial fields
$e^\er_{AB}$, $f_{AB}$, $\xi_{ABCD}$, $\chi_{(AB)CD}$, $\Theta_{ABCD}$
and $\phi_{ABCD}$ on the initial hypersurface $\mathcal{S}$. This
information can be calculated from a solution to the conformal
Hamiltonian and momentum constraints.

\medskip
If the data are close to de Sitter data, then on $\mathcal{S}$
(i.e. $\tau=0$) one has that
\begin{subequations}
\begin{eqnarray}
&& e^0_{AB}=0 \label{StdSdata1}\\
&& e^\er_{AB} =\sigma^\er_{AB}, \\
&& f_{AB}= 0, \\
&& \xi_{ABCD}= -\frac{\mbox{i}}{2}h_{ABCD} + \breve{\xi}_{ABCD},\\
&& \chi_{(AB)CD}= \breve{\chi}_{(AB)CD}, \\
&& \Theta_{ABCD} = -\frac{1}{2}h_{ABCD} + \breve{\Theta}_{ABCD}, \\
&& \phi_{ABCD} = \breve{\phi}_{ABCD} \label{StdSdata7},
\end{eqnarray}
\end{subequations}
where quantities with a $\;\;\breve{}\;\;$ denote quantities which vanish for exactly de Sitter data.

\medskip Following the discussion in sections \ref{section:deSitter}
and \ref{section:standard:deSitter:data}, initial data for the
congruence of conformal geodesics will be chosen such that $\Theta_*=1
$, $b_*=0$, and hence $\dot{\Theta}_* = \langle d,v \rangle_*=0$. Upon
this choice of data for the congruence, one has that the location of
the conformal boundary of the development is given by (\ref{taupmdS1})
to be $\tau =\pm 2$ as in the case of the de Sitter space-time. Finally,
the data (\ref{StdSdata1})-(\ref{StdSdata7}) is supplemented by data
for the Jacobi field $\eta_{AA'}$ of the form
\begin{equation}
\label{Jacobi:data}
\eta=0, \quad \eta_{AB}^\er= \sigma^\er_{AB}.
\end{equation}

\bigskip
From the previous discussion and theorem
\ref{thm:mod:Kato} one has the following existence and stability
result.

\begin{theorem}
\label{thm:standard:deSitter}
  Suppose $m\geq 4$. Let $u_0 = u'_0 + \breve{u}_0$ be standard de
  Sitter-like Cauchy initial data. There exists $\varepsilon>0$ such
  that if $\parallel \breve{u}_0 \parallel_m <\varepsilon $ then there
  is a unique solution $u'+\breve{u}$ to the conformal propagation
  equations (\ref{p1})-(\ref{p7}) and (\ref{b0})-(\ref{b4}) with
  minimal existence interval $\tau\in[-2,2]$ with $u\in
  C^{m-2}([-2,2]\times \mathcal{S})$ and such that the associated congruence of
  conformal geodesics contains no conjugate points in $[-2,2]$. The
  fields $u'+\breve{u}$ imply a $C^{m-2}$ solution to the vacuum
  Einstein field equations with positive cosmological constant for
  which the sets $\mathscr{I}^\pm =\{\pm 2\} \times \mathcal{S}$
  represent future and past conformal infinity.
\end{theorem}

\textbf{Proof.} Local existence to the system of the form
(\ref{sym:hyp:system}) implied by the propagation equations
(\ref{p1})-(\ref{p7}), (\ref{b0})-(\ref{b4}) and (\ref{j1})-(\ref{j2})
follows from point (i) in theorem \ref{thm:mod:Kato} and the
observation that if $\varepsilon>0$ is suitably small then
$\breve{u}\in D^m_\delta$. The reference solution given by
(\ref{Ecosmos1})-(\ref{Ecosmos7}) has existence interval
$(-\infty,\infty)\supset[-2,2]$, and the Jacobi fields associated to
the congruence of conformal geodesics never vanish. Thus, from (ii)
and (iii) in theorem \ref{thm:mod:Kato} one has that for suitably
small $\varepsilon>0$ the developments of all data such that
$\parallel \breve{u}_0 \parallel_m <\varepsilon $ have a minimum
existence interval $[-2,2]$. By reducing $\varepsilon$, if necessary,
one can ensure that $\eta_{AB}\neq 0$ for $\tau\in [-2,2]$ so that no
conjugate points arise. \hfill $\Box$

\subsubsection{Asymptotic Cauchy data for de Sitter-like space-times}
In the case of asymptotic Cauchy data for de Sitter-like space-times,
whereby information is prescribed on an initial hypersurface $\mathcal{S}$ which will be regarded as the past conformal infinity of the development, one
has that the initial value of the fields $e^\er_{AB}$, $f_{AB}$,
$\xi_{ABCD}$, $\chi_{(AB)CD}$, $\Theta_{ABCD}$ is calculated from the
following fields on $\mathscr{I}^-$: a 3-metric $h_{\alpha\beta}$,
$\updn{\chi}{\alpha}{\alpha}$ and a symmetric trace-free tensor field,
$d_{\alpha\beta}$, satisfying $D^\alpha d_{\alpha\beta}=0$, where $D$
denotes the Levi-Civita connection of $h_{\alpha\beta}$. The form of
the data for the propagation equations is formally identical to that
of the data (\ref{StdSdata1})-(\ref{StdSdata7}).

\medskip Following the discussion of sections \ref{section:deSitter}
and \ref{section:asymptotic:deSitter:data} initial data for the
congruence of conformal geodesics is chosen, without loss of
generality, such that $\langle d, v \rangle_*=1$ and
$\ddot{\Theta}_*=-1/2$. The data for the Jacobi field is taken to be
identical to that in equation (\ref{Jacobi:data}).

\medskip
The corresponding existence, uniqueness and stability result for this case is the following.

\begin{theorem}
\label{thm:asymptotic:deSitter}
 Suppose $m\geq 4$. Let $u_0 = u'_0 + \breve{u}_0$ be asymptotic de
  Sitter-like Cauchy initial data. There exists $\varepsilon>0$ such
  that if $\parallel \breve{u}_0 \parallel_m <\varepsilon $ then there
  is a unique solution $u'+\breve{u}$ to the conformal propagation
  equations (\ref{p1})-(\ref{p7}) and (\ref{b0})-(\ref{b4}) with
  minimal existence interval $\tau\in[0,4]$ with $u\in
  C^{m-2}([0,4]\times \mathcal{S})$ and such that the associated congruence of
  conformal geodesics contains no conjugate points in $[0,4]$. The
  fields $u'+\breve{u}$ imply a $C^{m-2}$ solution to the vacuum
  Einstein field equations with positive cosmological constant for
  which the sets $\mathscr{I}^- =\{ 0 \} \times \mathcal{S}$ and $\mathscr{I}^+=\{ 4 \} \times \mathcal{S}$ represent, respectively, past and future conformal
  infinity.
\end{theorem}

\textbf{Proof.} The proof is identical to that of theorem
\ref{thm:standard:deSitter}.

\subsection{Application to Minkowski-like space-times}
\label{section:thm:Minkowski}
In the case of hyperboloidal initial data one starts with a solution
$(\Omega,\Sigma, h_{\alpha\beta}, \chi_{\alpha\beta})$ to the
$\lambda=0$ conformal Hamiltonian and momentum constraints:
\begin{eqnarray*}
&& 2\Omega D_\alpha D^\alpha \Omega -3 D_\alpha\Omega D^\alpha \Omega +\frac{1}{2} \Omega^2r -3 \Sigma^2 -\frac{1}{2} \Omega^2 \left( \chi^2 -\chi_{\alpha\beta}\chi^{\alpha\beta}\right) + 2 \Omega \Sigma \chi=0, \nonumber \\
&& \Omega^3 D^{\alpha}(\Omega^{-2}\chi_{\alpha\beta})-\Omega \left( D_\beta \chi -2 \Omega^{-1}D_\beta \Sigma \right)=0,  \nonumber
\end{eqnarray*}
where $r$ denotes the Ricci scalar of the metric $h_{\alpha\beta}$ and
$\chi=h^{\alpha\beta}\chi_{\alpha\beta}$. The solution to the
conformal Hamiltonian and momentum constraints satisfies
\begin{eqnarray*}
\Omega=0, \quad \Sigma <0, \quad h^\sharp(D\Omega,D\Omega) = -\Sigma^2  &&\mbox{ on } \mathcal{Z}\equiv \partial \overline{\mathcal{S}}, \nonumber \\
 \Omega>0 &&\mbox{ on } \overline{\mathcal{S}}\setminus \mathcal{Z}.\nonumber
\end{eqnarray*}
The value of the fields $\xi_{ABCD}$, $\chi_{(AB)CD}$ are calculated
from $(\Omega,\Sigma, h_{\alpha\beta}, \chi_{\alpha\beta})$. We set
$\Theta_*=\Omega $, $\dot{\Theta}_* = \Sigma$ and $e^\er_{AB}$,
$f_{AB}$ as outlined before. The initial values of $\Theta_{ABCD}$ and
$\phi_{ABCD}$ on the initial hyperboloid $\overline{\mathcal{S}}$ are
then calculated from the above conformal constraint equations.

\medskip Theorem \ref{thm:mod:Kato} gives existence for symmetric
hyperbolic systems of the form (\ref{sym:hyp:system}) with data
prescribed on an initial manifold, $\mathcal{S}$, which is
topologically $\Sphere^3$. Consequently the data on
$\overline{\mathcal{S}}$ has to be extended to data on the whole of
$\mathcal{S}$. Noting that $\mathcal{S}$ is diffeomorphic to
$\Sphere^3$, we consider in what follows $\overline{\mathcal{S}}$ as a
subset of $\Sphere^3$. As discussed in \cite{Fri86b}, there is a
linear extension operator
$E:H^m(\overline{\mathcal{S}},\Real^N)\longrightarrow
H^m(\Sphere^3,\Real^N)$ such that if $v\in
H^m(\overline{\mathcal{S}},\Real^N)$ then $(Ev)(x)=v(x)$ almost
everywhere in $\overline{\mathcal{S}}$ and $\parallel E v
\parallel_{m} \leq K \parallel v
\parallel_{m,\overline{\mathcal{S}}}$, with $K$ a constant which is
universal for fixed $m$. As in the cases of the de Sitter-like
space-times, the data for the equation (\ref{sym:hyp:system}) takes
the form $u_0=u'_0 + \breve{u}_0$. The vector $u'_0$ is defined as in
equations (\ref{StdSdata1})-(\ref{StdSdata7}) and thus it is defined
on the whole of $\mathcal{S}$. On the other hand, the vector
$\breve{u}_0$ is only defined on $\overline{\mathcal{S}}$, and then
needs to be extended. We define the \emph{extended data}
$\mathring{u}_0$ by
\[
\mathring{u}_0 = u'_0 + E\breve{u}_0.
\]
It should be mentioned that the extension of the data is, in
principle, non-unique and that in general $\mathring{u}_0$ will not
satisfy the conformal constraint equations on $\mathcal{S}\setminus
\overline{\mathcal{S}}$. This will not have an effect on the
development of the hyperboloidal data as 
$D^+(\overline{\mathcal{S}}) \cap I^+(\mathcal{S} \setminus \overline{\mathcal{S}}) = \emptyset$.

\medskip In the case of Minkowski-like data we shall make use of a
conformal factor $\Theta$ of the form given by formula
(\ref{massaged_Theta}) with $\alpha < 0$. For a sufficiently small
ball of data centred on Minkowski data this implies that the time
$\tau_{i^+}$ is close to the one of Minkowski, namely $\tau_{i^+_M} =
2$.  The spatial location of $i^+$ is given by the condition $D_k
\Omega = 0 $ and will be close to that of $i^+_M$.

\bigskip The existence and stability result for Minkowski-like
hyperboloidal data is the following.

\begin{theorem}
\label{thm:Minkowski}
 Suppose $m\geq 4$. Let $u_0 = u'_0 + \breve{u}_0$ be Minkowski-like
initial data. Given $T_0 > 2 $ there exists $\varepsilon>0$ such that 
\begin{enumerate}
\item[(i)] for $\parallel \breve{u}_0 \parallel_m <\varepsilon $ then there is a solution
$u'+\breve{u}$ to the conformal propagation equations (\ref{p1})-(\ref{p7})
 and (\ref{b0})-(\ref{b4}) with minimal existence interval $\tau \in [0,T_0]$ 
 and $u\in C^{m-2}([0,T_0] \times \mathcal{S})$;
\item[(ii)] the associated congruence of conformal geodesics contains no conjugate points in $[0,T_0]$;
\item[(iii)] for every $\breve{u}_0 \in B_\varepsilon(0)$ there is a unique point in 
$\overline{\mathcal{S}}\setminus \mathcal{Z} $ such that $D_k \Omega = 0 $;
\item[(iv)] for all $\breve{u}_0 \in B_\varepsilon(0)$ we have $\tau_{i^+} \in [0,T_0] $.
\end{enumerate}
The solution $u'+\breve{u}$ is unique on
$\mathcal{D}^+(\overline{\mathcal{S}})$, the domain of dependence of
$\overline{\mathcal{S}}$ and implies a $C^{m-2}$ solution to the
vacuum Einstein field equations with vanishing cosmological constant
for which the set $\mathscr{I}^+$ as given by (\ref{scri:Minkowski})
represents future null infinity while the point $i^+$ given by the
conditions $\tau = \tau_{i^+}$ and $D_k\Omega=0$ represents timelike
infinity.
\end{theorem}

\textbf{Proof.} As before, local existence to the system of the
form (\ref{sym:hyp:system}) implied by the propagation equations
(\ref{p1})-(\ref{p7}) and (\ref{b0})-(\ref{b4}) follows from point
(i) in theorem \ref{thm:mod:Kato} and the observation that if
$\varepsilon>0$ is suitably small then $u'_0+ E\breve{u}_0\in
D^m_\delta$. The reference solution given by
(\ref{Ecosmos1})-(\ref{Ecosmos7}) has existence interval
$(-\infty,\infty)\supset[0,T_0]\supset[0,-2/\alpha]$, and the Jacobi fields
associated to the congruence of conformal geodesics never vanish.
Note that $T_0$ is chosen independently of the initial data.
Thus, from (ii) and (iii) in theorem \ref{thm:mod:Kato} one has
that for suitably small $\varepsilon>0$ the developments of all
extended data such that $\parallel E\breve{u}_0 \parallel_m
<\varepsilon/K $ have a minimum existence interval
$[0,T_0]$. By reducing $\varepsilon$, if necessary, one
can ensure that $\eta_{AB}\neq 0$ for $\tau\in [0,T_0]$ so
that no conjugate points arise on $\mathcal{S}\times
[0,T_0]$. In particular, this also holds for
$\mathcal{D}^+(\overline{\mathcal{S}})\subset \mathcal{S}\times
[0,T_0]$. The characterisation of the conformal boundary
then follows from the discussion of section
\ref{section:Minkowski:data}\hfill $\Box$

\section{Conclusion and remarks}
\label{section:conclusions}

In this article we reinvestigated the problem of de Sitter-like and
Minkowski-like space-times with the help of their conformal structure.
The use of conformal Gaussian coordinates has several advantages.
Their construction is conformally invariant and provides a natural
conformal factor along the congruence, which for vacuum can be
calculated explicitly. Furthermore, the associated Weyl connection
yields a gauge in which the extended conformal field equations are
simplified and a symmetric hyperbolic system is obtained. It was shown
that for these vacuum space-times the location of the conformal
boundary $\mathscr{I}^+ \cup i^+$ can be read off directly from the
initial data. It should be mentioned that this formulation of the
initial value problem for the conformal Einstein field equations is
amenable to numerical implementations, and indeed, a frame version
thereof has been used in the numerical investigations of cosmological
space-times described in \cite{Bey08,Bey09}.

For our calculations certain choices of initial data were motivated by
the behaviour of the exact solution and their conformal embedding into
the Einstein cylinder. Hence on the initial surface we set
$b_k=\Upsilon_k$. It should be noted that not all congruences of
conformal geodesics are suitable as they can develop conjugate points
before or at the conformal boundary. A simple example is provided by
the standard time-like geodesics in Minkowski space, which are
conformal geodesics with $b_k=0$. In the conformally compactified
picture they converge at $i^+$, where the congruence has a conformal
conjugate point. This is the only point of the conformal boundary that
is reached. Moreover, it takes infinite time $\tau$ to get there and
the induced conformal factor is constant along each curve.

Other choices of initial data are related to the parametrisation of
the congruence, e.g. for de Sitter-like space-times $\ddot{\Theta}_* $
is free datum on $\mathscr{I}^- $ . These choices affect the location
of $\mathscr{I}^+ $ in our conformal Gaussian coordinates and as long
as we avoid $\tau \to \infty$ before reaching $\mathscr{I}^+ $ we can
specify them suitably. However, these choices do not reduce the class
of de Sitter-like space-times for which theorem
\ref{thm:asymptotic:deSitter} holds.

Finally, it is mentioned that the methods discussed here can be
adapted to discuss the stability of other suitable reference solutions
like the \emph{purely radiative space-times} of \cite{Fri86b}. The
detailed discussion of this generalisation will be presented elsewhere.

\section*{Acknowledgements}
CL would like to thank the Leverhulme Trust for a research project
grant (F/07 476/AI). JAVK is an EPSRC Advanced Research Fellow. JAVK
would like to thank the hospitality of the Mittag-Leffler Institute in
Djursholm-Stockholm, Sweden, where parts of this project were carried
out as part of the programme \emph{Geometry, Analysis and General
  Relativity}. We are thankful to CM Losert-VK for a careful reading
of the manuscript.

\appendix
\section{An alternative discussion of  the conjugate points}
\label{appendix:conjugate:points}

In the main part of the article 
the variable $u=(\upsilon, \phi)$ contained the Jacobi fields and thus
guaranteed that one can avoid conjugate points. In this appendix we
outline an alternative approach. Kato's theorem ---cfr. theorem
\ref{thm:mod:Kato}--- can be used for an unknown $u$ which
does not contain $\eta_{AB}$. It states that one can find initial data
close to the original data such that the components of the curvature
tensors stay with in certain bounds. This approach will be taken here.
The Jacobi fields are here treated outside the symmetric hyperbolic
system following work in \cite{TodLue07}.

Since the reference solution is conformally flat its Weyl tensor and Cotton-York tensor vanish.  However the second
fundamental form and the Schouten tensor do not vanish and hence it is
more difficult to obtain suitable bounds for the estimations that are
to follow. Hence we use the second order Jacobi
equation and replace the Schouten tensor with the Cotton-York tensor in
it.

\noindent The conformal Jacobi equation (\ref{Jacobi2}) is linear. Thus it
is sufficient to consider initial data with
\[
z(0)=1, \quad \eta^\mu(0)=e^\mu_k,\quad \dot{z}(0)=\chi_{kk}. 
\]
Furthermore, let $X\equiv \dot{z}(0)$. We split the Jacobi fields into 
\[
\eta^\mu = \eta_0 v^\mu + z m^\mu
\]
where $z^2\equiv -h(\eta,\eta)$ and $h(m,m)=-1 $. Denoting $\mathrm{d}/\mathrm{d}\tau $ by $D$, we can derive
\[ 
\ddot{z} = - \frac{h(D^2 \eta, \eta)}{z}+\frac{h(D\eta, D\eta)h(\eta, \eta) - (h(D \eta, \eta))^2}{z^3}  \ge  - \frac{h(D^2 \eta, \eta)}{z} = (E_{ik} + \hat{L}_{ik})m^i m^k z,
\]
where we have used the Cauchy-Schwartz inequality and the conformal
Jacobi equation (\ref{Jacobi2}).

\noindent We observe the identity 
$
D(\hat{L}(\eta,e_k)) = \hat{Y}(v,\eta,e_k)
$
and its integral form
\[ \int^\tau_0 \hat{Y}_{0jk}\eta^j \mbox{d}\sigma = \hat{L}(\eta,e_k)(\tau) - \hat{L}(\eta,e_k)(0).
\]
This allows us to replace the Schouten tensor in the above inequality
and obtain
\[  \ddot{z} \ge E(m,m)(\tau)z(\tau) + \int^\tau_0 (\hat{Y}_{0jk}zm^j)(\sigma)m^k(\tau) \mbox{d}\sigma  + (z \hat{L}_{ik}m^i)(0) m^k(\tau) . 
\]

Observing that $m$ is a space-like unit vector and employing a
Cauchy-Schwartz type argument in combination with the existence and
stability theorem (\ref{thm:mod:Kato}), we can see that there exist
bounds such that
\[
-K \le E(m,m)(\tau) \le K \quad 
-L \le \hat{Y}_{0jk}m^j(\sigma)m^k(\tau) \le L.
\] 
For the Schouten tensor on the initial surface we use 
\[
-Q \le
\hat{L}_{ik}z\,m^i(0) m^k(\tau) \le Q',
\]
with $Q, Q' \ge 0$. This simplifies the inequality to give
\[
\ddot{z} \ge -K z(\tau)  -L \int^\tau_0 z(\sigma) d\sigma  - Q. 
\]

One now follows the argument of \cite{TodLue07}, by introducing a family
of functions $y_\varepsilon(\tau) $ satisfying
\begin{equation}
\label{yODE}
\ddot{y_\varepsilon} = -(K + \varepsilon^2) y_\varepsilon(\tau)  -L \int^\tau_0 y_\varepsilon(\sigma) d\sigma  - Q
\end{equation}
with initial data $y_\varepsilon(0)=1$, $\dot{y}_\varepsilon(0)=X$.
The idea is to show that $z(\tau)$ cannot vanish before
$y_\varepsilon(\tau) $, in particular when $\varepsilon=0 $. By
estimating a lower bound for the time $\tau_y$ when
$y_{\varepsilon=0}$ vanishes, we get a lower bound for the time before
the solution can develop conjugate points.

We briefly recall the steps in the argument, whose details can be
found in \cite{TodLue07,Cla78}. We drop
the subscript $\varepsilon$. One defines $R\equiv z/y$ and
$W\equiv\dot{z}y - \dot{y}z = y^2 \dot{R}$. From this one derives
$$
  \dot{W}(\tau) \ge L \int^\tau_0 y(\tau)y(\sigma)  [R(\tau)-R(\sigma)] d\sigma + Q y(\tau)[R(\tau)-1] .
 $$
 There must be an interval $J=[0,T_\varepsilon]$ such that $W \ge 0$
 and hence that $R\ge 1$ and $\dot{W} > 0 $ hold on $J$. This implies
 that as long as $y$ does not vanish, we can extend $J$ further while
 $R \ge 1$ holds. Hence $z \ge y \ge 0$ on $J$. By continuity in
 $\varepsilon$, $\lim_{\varepsilon \to 0} T_\varepsilon = T_0 >0 $.
 Thus we only need to focus on $y=y_{\varepsilon=0}$, for which
 integrate (\ref{yODE}) twice to get
\[
 y(\tau) = 1 - X\tau - \frac{1}{2} Q\tau^2 - K  \int^\tau_0 (\tau-\sigma)y(\sigma)d\sigma -  L  \int^\tau_0 \frac{1}{2}(\tau-\sigma)^2 y(\sigma)d\sigma.
\]
Observing that $y(\tau) \le y(0) + X \tau$ and using it to obtain
upper bounds for the integrals we get
\[
y(\tau) \ge Y(\tau)\equiv 1 - X\tau - \frac{1}{2} (Q+K)\tau^2 - \frac{1}{6}(XK+L)\tau^3 - \frac{LX}{24}\tau^4
\]
For the reference solution all constants vanish so that $z \ge y \ge
1$ for all $\tau$, which agrees with $z=1+\tau^2/4$. If one of the
constants $X,Q,K,L$ is non-zero, we get exactly one positive root, as
all points of inflection have to lie in the second and third quadrant.

Without loss of generality we can set all of them to $R$ by appealing
to Kato's theorem. Suppose we now fix $\tau=T$ suitably beyond the
values at future infinity of the space-time we would like to perturb.
Then $F(T)=1-A(T)R - B(T)R^2$, for some constants $A(T)$, $B(T)$. It is
clear that we can always choose the perturbation of the initial data
suitably small in Kato's theorem, so that the bound $R$ guarantees
that $f(T)>0$ and hence that the perturbed solution will not develop
conjugate points before the chosen time $\tau=T$.


\end{document}